\documentclass[twocolumn, twocolappendix]{aastex63}

\usepackage{rotating}
\usepackage{amsmath}
\usepackage{mathrsfs}
\usepackage{afterpage}
\usepackage{scalerel}
\usepackage{natbib}
\bibpunct[; ]{(}{)}{;}{a}{}{,}
\usepackage{multirow}
\usepackage{graphics}
\usepackage{threeparttable}
\usepackage[varg]{txfonts}
\usepackage{xcolor}
\usepackage{bm}
\usepackage{wrapfig}
\hypersetup{colorlinks,linkcolor={blue},citecolor={blue},urlcolor={blue}}

\begin{document}
\title{Mapping the Growth of Supermassive Black Holes as a Function of Galaxy Stellar Mass and Redshift}

\author[0000-0002-4436-6923]{Fan Zou}
\affiliation{Department of Astronomy and Astrophysics, 525 Davey Lab, The Pennsylvania State University, University Park, PA 16802, USA}
\affiliation{Institute for Gravitation and the Cosmos, The Pennsylvania State University, University Park, PA 16802, USA}

\author[0000-0002-6990-9058]{Zhibo Yu}
\affiliation{Department of Astronomy and Astrophysics, 525 Davey Lab, The Pennsylvania State University, University Park, PA 16802, USA}
\affiliation{Institute for Gravitation and the Cosmos, The Pennsylvania State University, University Park, PA 16802, USA}

\author[0000-0002-0167-2453]{W. N. Brandt}
\affiliation{Department of Astronomy and Astrophysics, 525 Davey Lab, The Pennsylvania State University, University Park, PA 16802, USA}
\affiliation{Institute for Gravitation and the Cosmos, The Pennsylvania State University, University Park, PA 16802, USA}
\affiliation{Department of Physics, 104 Davey Laboratory, The Pennsylvania State University, University Park, PA 16802, USA}

\author[0000-0003-0334-8742]{Hyungsuk Tak}
\affiliation{Department of Statistics, The Pennsylvania State University, University Park, PA 16802, USA}
\affiliation{Department of Astronomy and Astrophysics, 525 Davey Lab, The Pennsylvania State University, University Park, PA 16802, USA}
\affiliation{Institute for Computational and Data Sciences, The Pennsylvania State University, University Park, PA 16802, USA}

\author[0000-0001-8835-7722]{Guang Yang}
\affiliation{Kapteyn Astronomical Institute, University of Groningen, P.O. Box 800, 9700 AV Groningen, The Netherlands}
\affiliation{SRON Netherlands Institute for Space Research, Postbus 800, 9700 AV Groningen, The Netherlands}

\author[0000-0002-8577-2717]{Qingling Ni}
\affiliation{Max-Planck-Institut f\"{u}r extraterrestrische Physik (MPE), Gie{\ss}enbachstra{\ss}e 1, D-85748 Garching bei M\"unchen, Germany}

\email{E-mail: fanzou01@gmail.com}

\begin{abstract}
The growth of supermassive black holes is strongly linked to their galaxies. It has been shown that the population mean black-hole accretion rate ($\overline{\mathrm{BHAR}}$) primarily correlates with the galaxy stellar mass ($M_\star$) and redshift for the general galaxy population. This work aims to provide the best measurements of $\overline{\mathrm{BHAR}}$ as a function of $M_\star$ and redshift over ranges of $10^{9.5}<M_\star<10^{12}~M_\odot$ and $z<4$. We compile an unprecedentedly large sample with eight thousand active galactic nuclei (AGNs) and 1.3 million normal galaxies from nine high-quality survey fields following a wedding-cake design. We further develop a semiparametric Bayesian method that can reasonably estimate $\overline{\mathrm{BHAR}}$ and the corresponding uncertainties, even for sparsely populated regions in the parameter space. $\overline{\mathrm{BHAR}}$ is constrained by \mbox{X-ray} surveys sampling the AGN accretion power and UV-to-infrared multi-wavelength surveys sampling the galaxy population. Our results can independently predict the \mbox{X-ray} luminosity function (XLF) from the galaxy stellar mass function (SMF), and the prediction is consistent with the observed XLF. We also try adding external constraints from the observed SMF and XLF. We further measure $\overline{\mathrm{BHAR}}$ for star-forming and quiescent galaxies and show that star-forming $\overline{\mathrm{BHAR}}$ is generally larger than or at least comparable to the quiescent $\overline{\mathrm{BHAR}}$.
\end{abstract}

\section{Introduction}
\label{sec: intro}
Supermassive black holes (SMBHs) and galaxies appear to be fundamentally linked (e.g., \citealt{Kormendy13}). Especially, the SMBH mass ($M_\mathrm{BH}$) and the galaxy bulge mass are found to be tightly correlated in the local universe, and the cosmic evolution of the SMBH accretion density and the star-formation density are broadly similar and both peak at $z\approx2$. Therefore, it is a fundamental question in extragalactic astronomy to understand how cosmic SMBH growth depends on galaxy properties.\par
SMBHs grow primarily through rapid accretion when they are observed as active galactic nuclei (AGNs); mergers are an additional growth mode. \mbox{X-ray} emission is known to be a good indicator of AGN activity because of its universality among AGNs, high penetrating power through obscuration, and low dilution from galaxy starlight (e.g., \citealt{Brandt15, Brandt22}). Therefore, \mbox{X-ray} surveys can be used to constrain the accretion distribution and the black-hole accretion rate ($\mathrm{BHAR}=dM_\mathrm{BH}/dt$) of SMBHs (e.g., \citealt{Aird12, Aird18, Bongiorno12, Bongiorno16, Georgakakis17, Wang17, Yang17, Yang18a, Yang19, Ni19, Ni21b}). However, the duration of a galaxy within the AGN phase is relatively short, and AGNs also have strong variability (e.g., \citealt{Hickox14, Yang16}); thus, BHAR is often averaged over a large galaxy sample as an estimator of $\overline{\mathrm{BHAR}}$, the long-term population mean BHAR.\par
$\overline{\mathrm{BHAR}}$ has been shown to be redshift-dependent and correlated with both stellar mass ($M_\star$) and star-formation rate (SFR), while the $M_\star$ dependence is more fundamental for the general galaxy population (e.g., \citealt{Hickox14, Yang17, Yang18a}). In recent years, it has been found that $\overline{\mathrm{BHAR}}$ is also strongly related to galaxy morphology (e.g., \citealt{Ni19, Ni21b, Yang19, Aird22}), which may be more fundamental than the $\overline{\mathrm{BHAR}}-M_\star$ relation. However, such morphological measurements are expensive to obtain and require superb image resolution from, e.g., the Hubble Space Telescope, which inevitably are limited to small sky areas and thus can only provide a limited sample size covering a limited parameter space. In contrast, $M_\star$ and SFR are much more accessible. Notably, modern multi-wavelength photometric surveys have provided extensive photometric data, based on which $M_\star$ and SFR can be estimated by fitting the corresponding spectral energy distributions (SEDs; e.g., \citealt{Zou22}). Therefore, a well-measured $\overline{\mathrm{BHAR}}-M_\star$ relation is still essential to link SMBHs and galaxies.\par
The latest version of $\overline{\mathrm{BHAR}}$ as a function of $(M_\star, z)$ was derived in \citet{Yang18a} using the data from the Cosmic Evolution Survey (COSMOS; \citealt{Laigle16, Weaver22}), Great Observatories Origins Deep Survey (GOODS)-S, and GOODS-N \citep{Grogin11, Koekemoer11}. Although the relation in \citet{Yang18a} has proved to be successful over the years, there are still some limitations. First, although COSMOS, GOODS-S, and GOODS-N are sufficiently deep to probe $\overline{\mathrm{BHAR}}$ up to $z\approx4$, they cannot effectively sample the last half of cosmic time ($z\lesssim0.8$) because of their limited sky areas. Second, \citet{Yang18a} adopted strong assumptions when parametrically estimating $\overline{\mathrm{BHAR}}$, which may lead to underestimated $\overline{\mathrm{BHAR}}$ uncertainties. Built upon \citet{Yang18a}, this work aims to provide the best available map of $\overline{\mathrm{BHAR}}(M_\star, z)$ with the currently best data and statistical methodology. Now is indeed the right time to re-measure $\overline{\mathrm{BHAR}}$ given the fact that the past five years have witnessed the completion of several large, sensitive \mbox{X-ray} surveys in fields together with deep optical-to-IR surveys (e.g., \citealt{Chen18, Ni21}). These new X-ray surveys, when combined with previous ones, can return a large AGN sample over ten times larger than previous ones, as will be discussed in Section~\ref{sec: data}. In this work, we compile an unprecedentedly large sample from nine well-studied survey fields, many of which were finished after \citet{Yang18a} and even within $\lesssim2$~yrs before this work. Our adopted surveys follow a wedding-cake design and contain both deep, pencil-beam and shallow, wide ones, allowing us to effectively explore a wide range of parameter space. We further develop a semiparametric Bayesian approach that can return reasonable estimations and uncertainties, even for sparsely populated regions in the parameter space.\par
This work is structured as follows. Section~\ref{sec: data} describes the data. Section~\ref{sec: modeling} presents our methodology and $\overline{\mathrm{BHAR}}$ measurements. In Section~\ref{sec: discussion}, we discuss the implications of our results. Section~\ref{sec: summary} summarizes this work. We adopt a flat $\Lambda\mathrm{CDM}$ cosmology with $H_0=70~\mathrm{km~s^{-1}~Mpc^{-1}}$, $\Omega_\Lambda=0.7$, and $\Omega_M=0.3$.

\section{Data and Sample}
\label{sec: data}
We use the data within the Cosmic Assembly Near-Infrared Deep Extragalactic Legacy Survey (CANDELS) fields, four of the Vera C. Rubin Observatory Legacy Survey of Space and Time (LSST) Deep-Drilling Fields (DDFs), and eROSITA Final Equatorial Depth Survey (eFEDS) field. CANDELS and the LSST DDFs both contain several distinct fields, and we put those individual fields sharing similar areas and depths under the same umbrella (``CANDELS'' or ``LSST DDFs'') for convenience. Our adopted fields have \mbox{X-ray} coverage to provide AGN information and quality multi-wavelength data, which are essential for measuring galaxy properties. We summarize our field information in Table~\ref{tbl_data} and discuss them in the following subsections.\par

\begin{table*}
\begin{rotatetable*}
\caption{Basic Information for the Fields Used in This Work}
\label{tbl_data}
\centering
\begin{threeparttable}
\begin{tabular}{cccccccccc}
\hline
\hline
Field & Area & $m_\mathrm{lim}$ & \mbox{X-ray} Depth & \mbox{X-ray} ref. & Galaxy ref. & Photo-$z$ ref. & AGN & Galaxies & $(a, b)$\\
& ($\mathrm{deg^2}$) & (AB mag) & (ks)\\
(1) & (2) & (3) & (4) & (5) & (6) & (7) & (8) & (9) & (10)\\
\hline
GOODS-S & 0.05 & $26.5~(H)$ & 7000 (Chandra) & 5 & 1 & 4,8 & 224 (111) & 4144 & $(-15.87, 2.63)$\\
GOODS-N & 0.05 & $26.5~(H)$ & 2000 (Chandra) & 8 &1 & 1,11 & 174 (167) & 4603 & $(-15.49, 2.58)$\\
EGS & 0.06 & $26.5~(H)$ & 800 (Chandra) & 6 & 1 & 9 & 112 (10) & 5889 & $(-15.13, 3.08)$\\
UDS & 0.06 & $26.5~(H)$ & 600 (Chandra) & 4 & 1 & 8 & 117 (25) & 5010 & $(-15.05, 4.90)$\\
COSMOS & 1.27 & $24~(K_s)$ & 160 (Chandra) & 3 & 2 & 5,10 & 1459 (880) & 86765 & $(-14.68, 5.19)$\\
ELAIS-S1 & 2.93 & $23.5~(K_s)$ & 30 (XMM-Newton) & 7 & 3 & 6,12 & 676 (261) & 157791 & $(-13.90, 4.57)$\\
W-CDF-S & 4.23 & $23.5~(K_s)$ & 30 (XMM-Newton) & 7 & 3 & 6,12 & 872 (311) & 210727 & $(-13.86, 4.97)$\\
XMM-LSS & 4.20 & $23.5~(K_s)$ & 40 (XMM-Newton) & 2 & 3 & 2 & 1765 (898) & 254687 & $(-14.09, 5.36)$\\
eFEDS & 59.75 & $22~(Z)$ & 2 (eROSITA) & 1 & 2 & 3,7 & 2667 (1156) & 615068 & $(-13.51, 2.59)$\\
\hline
\hline
\end{tabular}
\begin{tablenotes}
\item
\textit{Notes.} (1) Field names. GOODS-S, GOODS-N, EGS, and UDS belong to CANDELS and are discussed in Section~\ref{sec: data_candels}. COSMOS, ELAIS-S1, W-CDF-S, and XMM-LSS belong to the LSST DDFs and are discussed in Section~\ref{sec: data_lsstddf}. eFEDS is discussed in Section~\ref{sec: data_efeds}. (2) Sky areas of the fields, only accounting for the regions we are using. (3) The limiting AB magnitudes we adopted in Section~\ref{sec: sample_construction} to calculate the $M_\star$ completeness curves, and the reference bands are written within parentheses. (4) The typical depths in exposure time of the \mbox{X-ray} surveys, and the parentheses list the observatories with which our adopted \mbox{X-ray} surveys were conducted. For XMM-Newton, the reported exposure is the typical flare-filtered one for a single EPIC camera. All the three EPIC cameras (one EPIC-pn and two EPIC-MOS) were used for the XMM-Newton observations, adding to $\approx80-100$~ks EPIC exposure in total. The ``$a$'' parameter values in Column~10 of this table represent typical flux limits in $2-10$~keV. (5) The references for the \mbox{X-ray} surveys. (6) The references for our adopted host-galaxy properties. All of these references have appropriately considered the AGN emission for AGNs. (7) Representative references examining the photo-$z$s in the corresponding fields. (8) Numbers of AGNs. The parentheses list the numbers of sources with spec-$z$s. The surface number density of eFEDS AGNs is much smaller than those in the other fields primarily because the eFEDS limiting magnitude is much brighter. (9) Numbers of normal galaxies. (10) The parameters describing the \mbox{X-ray} detection function; see Equation~\ref{eq: Pdet}. There is a subtle difference between eFEDS and other fields -- the eFEDS \mbox{X-ray} detection function is for the intrinsic $2-10$~keV flux, while the others' are for the observed flux.\\
\textit{\mbox{X-ray} references.} (1) \citet{Brunner22}; (2) \citet{Chen18}; (3) \citet{Civano16}; (4) \citet{Kocevski18}; (5) \citet{Luo17}; (6) \citet{Nandra15}; (7) \citet{Ni21}; (8) \citet{Xue16}.\\
\textit{Galaxy references.} (1) \citet{Yang19}; (2) \citet{Yu23}; (3) \citet{Zou22}.\\
\textit{Photo-$z$ references.} (1) \citet{Barro19}; (2) \citet{Chen18}; (3) \citet{Driver22}; (4) \citet{Luo17}; (5) \citet{Marchesi16}; (6) \citet{Ni21}; (7) \citet{Salvato22}; (8) \citet{Santini15}; (9) \citet{Stefanon17}; (10) \citet{Weaver22}; (11) \citet{Xue16}; (12) \citet{Zou21b}.
\end{tablenotes}
\end{threeparttable}
\end{rotatetable*}
\end{table*}

\subsection{CANDELS Fields}
\label{sec: data_candels}
CANDELS \citep{Grogin11, Koekemoer11} is a pencil-beam survey covering five fields -- GOODS-S ($0.05~\mathrm{deg^2}$), GOODS-N ($0.05~\mathrm{deg^2}$), Extended Groth Strip (EGS; $0.06~\mathrm{deg^2}$), UKIRT Infrared Deep Sky Survey Ultra-Deep Survey (UDS; $0.06~\mathrm{deg^2}$), and a tiny part of COSMOS (denoted as CANDELS-COSMOS hereafter; $0.06~\mathrm{deg^2}$). All the fields have ultra-deep UV-to-IR data (see, e.g., \citealt{Yang19} and references therein), allowing for detections of galaxies up to high redshift and low $M_\star$ and reliable measurements of these galaxies' properties. The first four have deep Chandra observations reaching $\sim$Ms depths from \citet[GOODS-S]{Luo17}, \citet[GOODS-N]{Xue16}, \citet[EGS]{Nandra15}, and \citet[UDS]{Kocevski18} and can thus effectively sample AGNs at high redshift and/or low luminosity. However, CANDELS-COSMOS shares the same \mbox{X-ray} depth as the full COSMOS field, and CANDELS-COSMOS is much smaller. Therefore, we do not use CANDELS-COSMOS but will instead directly analyze the full COSMOS field in Section~\ref{sec: data_lsstddf}.\par
We adopt the galaxy-property catalog in \citet{Yang19}, who have measured $M_\star$ and SFRs by fitting SEDs for all the CANDELS sources.\par

\subsection{LSST DDFs}
\label{sec: data_lsstddf}
The LSST DDFs (e.g., \citealt{Brandt18, Zou22}) include five fields -- COSMOS, Wide Chandra Deep Field-South (W-CDF-S), European Large-Area Infrared Space Observatory Survey-S1 (ELAIS-S1), XMM-Newton Large Scale Structure (XMM-LSS), and Euclid Deep Field-South (EDF-S). EDF-S has been selected as a LSST DDF only recently in 2022 and currently does not have sufficient data available, and we thus only use the former four original LSST DDFs with superb data accumulated over $\sim$ a decade. Note that this work does not use any actual LSST data because the Vera C. Rubin Observatory is still under construction at the time of writing this article.\par
COSMOS is a $\mathrm{deg^2}$-scale field with deep multi-wavelength data (e.g., \citealt{Weaver22}). \citet{Civano16} presented medium-depth ($\approx160$~ks) Chandra observations in COSMOS. The galaxy properties measured through SED fitting covering the \mbox{X-ray} to far-IR are taken from \citet{Yu23}. We only use the region with ``FLAG\_COMBINED = 0'' (i.e., within the UltraVISTA region and far from bright stars and image edges) in \citet{Weaver22} to ensure quality multi-wavelength characterizations. \citet{Ni21} presented $\approx30$~ks XMM-Newton observations in ELAIS-S1 and W-CDF-S, and \citet{Chen18} presented $\approx40$~ks XMM-Newton observations in XMM-LSS. The galaxy properties in these three fields are taken from \citet{Zou22}. We limit our analyses to the overlapping region between the \mbox{X-ray} catalogs and \citet{Zou22} because quality multi-wavelength data are essential for estimating photometric redshifts (photo-$z$s), $M_\star$, and SFRs. Besides, GOODS-S and UDS in Section~\ref{sec: data_candels} are embedded within W-CDF-S and XMM-LSS, respectively, and we remove the regions covered by GOODS-S and UDS to avoid double counting sources. Due to these reasons, our areas are slightly smaller than those reported in \citet{Chen18} and \citet{Ni21}.

\subsection{eFEDS}
\label{sec: data_efeds}
eFEDS is a $10^2~\mathrm{deg^2}$-scale field covered by eROSITA with $\approx2$~ks observations \citep{Brunner22}. We focus on the $60~\mathrm{deg^2}$ GAMA09 region \citep{Driver22} within eFEDS because the remaining parts do not have sufficient multi-wavelength data to constrain the host-galaxy properties (e.g., \citealt{Salvato22}). Unlike Chandra or XMM-Newton, eROSITA mostly works at $<2.3$~keV, which is more sensitive to obscuration. We thus rely on the \mbox{X-ray} properties cataloged in \citet{Liu22} for eFEDS sources, which are measured through detailed \mbox{X-ray} spectral fitting and thereby can largely overcome obscuration effects. As suggested in \citet{Liu22}, we only use sources with detection likelihoods $>10$ because fainter sources generally do not allow meaningful \mbox{X-ray} spectral analyses.

\subsection{Sample Construction}
\label{sec: sample_construction}
Sources in these fields all have either spectroscopic redshifts (spec-$z$s) or high-quality photo-$z$s, as have been extensively examined in previous literature. Representative examples are listed in Column~7 of Table~\ref{tbl_data}. Many more successful works built upon these redshifts have also indirectly justified their general reliability. When compared to the available spec-$z$s, the photo-$z$s are of high quality -- our sample has a $\sigma_\mathrm{NMAD}$ of 0.03 (0.04) and an outlier fraction of 4\% (15\%) for galaxies (AGNs).\footnote{Defining $\Delta z=z_\mathrm{phot}-z_\mathrm{spec}$, $\sigma_\mathrm{NMAD}$ is then the normalized median absolute deviation of $\Delta z/(1+z_\mathrm{spec})$, and outlier fraction is the fraction of sources with $\left|\Delta z\right|/(1+z_\mathrm{spec})>0.15$. These two parameters are standard metrics used to represent the photo-$z$ quality.} Spec-$z$s are adopted when available, and half of the involved AGNs have spec-$z$s.\par
We select sources with $0.05<z<4$ and $\left(\log M_{\star, \mathrm{min}}=9.5\right)<\log M_\star<\left(\log M_{\star, \mathrm{max}}=12\right)$. Sources labeled as stars are removed, as has been presented in the references in Column~6 of Table~\ref{tbl_data}. Only $\lesssim15\%$ of sources in each field are classified as stars. We apply a lower cut for $z$ because photo-$z$s are less reliable when too small (e.g., see Appendix~C of \citealt{Zou21b}), and the peculiar motions become non-negligible as well. We limit $\log M_\star>9.5$ because dwarf AGNs usually have much less reliable measurements and require special treatment (e.g., \citealt{Zou23}). We apply the same upper cuts as in \citet{Yang18a} for both $M_\star$ and $z$ because very few sources can exceed these thresholds.\par
We further construct a complete sample by applying redshift-dependent $M_\star$ cuts. To estimate the $M_\star$ depth for each field, we first adopt a reference band and denote its limiting magnitude as $m_\mathrm{lim}$. Following \citet{Pozzetti10}, we convert the magnitude depth to the expected limiting $M_\star$ for each galaxy with a magnitude of $m$: $\log M_\mathrm{lim}=\log M_\star+0.4(m-m_\mathrm{lim})$. At each redshift, we adopt the $M_\star$ completeness threshold as the value above which 90\% of the $M_\mathrm{lim}$ values lie. Sources below the $M_\star$ completeness curves are removed. For the CANDELS fields, we adopt the $H$ band with a limiting magnitude of 26.5~mag, and almost all the sources above our $\log M_\star$ cut of 9.5 are above the CANDELS $M_\star$ completeness curves, enabling constraints upon $\overline{\mathrm{BHAR}}$ in the low-$M_\star$ and high-$z$ regime. For the LSST DDFs, we adopt the $K_s$ band, and their limiting $K_s$ magnitudes are 24 for COSMOS \citep{Laigle16} and 23.5 for W-CDF-S, ELAIS-S1, and XMM-LSS \citep{Jarvis13}, respectively. For eFEDS, we adopt the $Z$ band with a limiting magnitude of 22. These $M_\star$ completeness cuts also automatically ensure the general SED-fitting reliability. The typical $i$-band magnitudes of sources at these limiting magnitudes are $i\approx24.8$ at $K_s=23.5$, $i\approx25.3$ at $K_s=24$, and $i\approx22.4$ at $Z=22$. These $i$-band magnitudes are roughly equal to the nominal ``depths'' of SEDs in \citet[see their Figure~30]{Zou22} and \citet{Yu23}, below which the number of available photometric bands may become small.\par
We then define $\lambda=L_\mathrm{X}/M_\star$, where $L_\mathrm{X}$ is the intrinsic \mbox{$2-10$~keV} luminosity, and we always adopt $\mathrm{erg~s^{-1}}~M_\odot^{-1}$ as the unit for $\lambda$. We use the \mbox{X-ray} surveys mentioned in the previous subsections to select AGNs. Following \citet{Aird12} and \citet{Yang18a}, we only use sources detected in the hard band (HB)\footnote{The detection energy range for the HB has slightly different definitions in different fields -- $2-7$~keV for CANDELS and COSMOS, $2-12$~keV for W-CDF-S and ELAIS-S1, and $2-10$~keV for XMM-LSS.} for CANDELS and the LSST DDFs. The reason is to minimize the effects of obscuration. Selecting AGNs in soft bands ($<2$~keV) is biased toward little or no absorption. Since the obscuration level is known to be correlated with $\lambda$ (e.g., \citealt{Ricci17}), soft-band-selected AGNs are expected to be biased in terms of $\lambda$. Besides, our analyses need intrinsic $L_\mathrm{X}$, and HB fluxes are the least affected by obscuration. To calculate $L_\mathrm{X}$ and, consequently, $\lambda$ of these HB-detected sources, we use Equation~A4 in \citet{Zou22} and adopt a photon index of 1.6. As discussed in \citet{Yang18a}, a photon index of 1.6 returns $L_\mathrm{X}$ agreeing the best with those from \mbox{X-ray} spectral fitting. For eFEDS, as mentioned in Section~\ref{sec: data_efeds}, we use the de-absorbed $0.5-2$~keV flux in \citet{Liu22} and convert it to $L_\mathrm{X}$ assuming a photon index of 1.8. Although eROSITA observations are more prone to obscuration effects, and it is less accurate to measure $L_\mathrm{X}$ with soft \mbox{X-rays} below $\approx2$~keV, we have verified in Appendix~\ref{append: efeds} that our median results remain similar when excluding eFEDS. It should be noted that we do not exclusively rely upon eFEDS to provide constraints at low-$z$ and/or high-$M_\star$. The LSST DDFs, especially with the \mbox{X-ray} coverage in \citet{Chen18} and \citet{Ni21} added, already have $12.6~\mathrm{deg^2}$ of coverage with useful HB sensitivity (see Table~\ref{tbl_data}), and thus can also provide beneficial constraints. We define AGNs as those with $\log\lambda>\log\lambda_\mathrm{min}=31.5$ and neglect the contribution of SMBHs with $\lambda\leq\lambda_\mathrm{min}$ to $\overline{\mathrm{BHAR}}$. This is because few of the \mbox{X-ray}-detected AGNs are below $\lambda_\mathrm{min}$, and the emission from \mbox{X-ray} binaries may become non-negligible for low-$\lambda$ sources. As we will show in Section~\ref{sec: bhar}, $\overline{\mathrm{BHAR}}$ is indeed dominated by sources above $\lambda_\mathrm{min}$.\par
\begin{figure*}
\centering
\resizebox{\hsize}{!}{\includegraphics{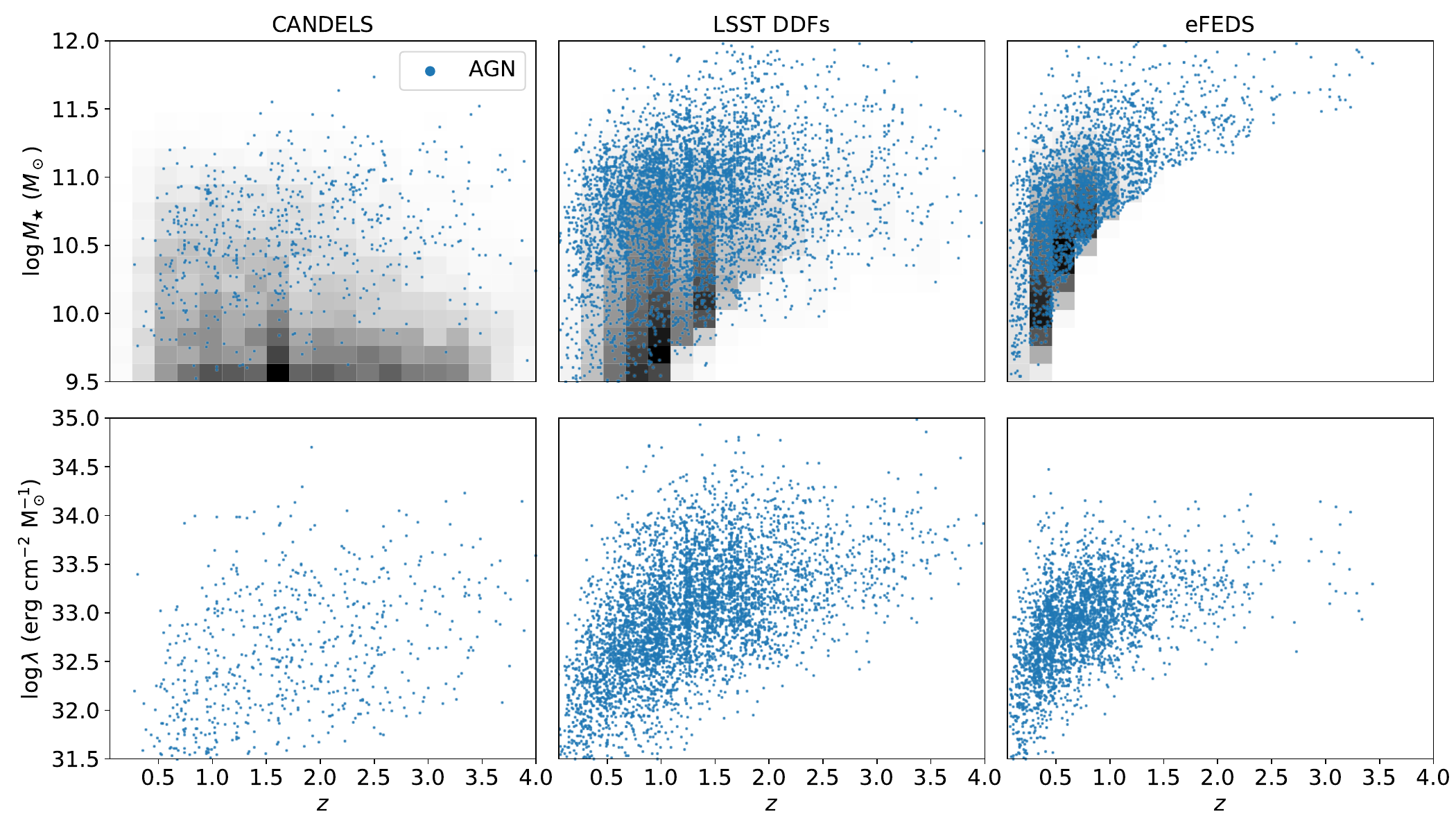}}
\caption{Our sample in the $z-M_\star$ (top) and $z-\lambda$ (bottom) planes. The left, middle, and right panels are for CANDELS, the LSST DDFs, and eFEDS, respectively. The points are AGNs. The background grey-scale cells in the left panel are the 2-D histogram of the number of normal galaxies, with darker cells representing more galaxies. The apparent deficiency of sources in the high-$z$ and/or low-$M_\star$ regime in the middle and right panels is due to our $M_\star$ completeness cuts.}
\label{fig_dataplot_total}
\end{figure*}

In total, we have eight thousand AGNs and 1.3 million normal galaxies, and they are plotted in the $z-M_\star$ and $z-\lambda$ planes in Figure~\ref{fig_dataplot_total}, where each column presents fields with comparable depths and areas. Note that \citet{Yang19}, \citet{Zou22}, and \citet{Yu23}, from which our adopted galaxy properties are taken, all have appropriately considered the AGN emission for AGNs. We will also assess the impact of AGNs that dominate the SEDs in Appendix~\ref{append: blagn}.

\section{Method and Results}
\label{sec: modeling}
Denoting $p(\lambda\mid M_\star, z)$ as the conditional probability density per unit $\log\lambda$ of a galaxy with $(M_\star, z)$ hosting an AGN with $\lambda$ and $k_\mathrm{bol}(L_\mathrm{X})$ as the $L_\mathrm{X}$-dependent $2-10$~keV bolometric correction (i.e., the ratio between the AGN bolometric luminosity and $L_\mathrm{X}$), $\overline{\mathrm{BHAR}}$ can be expressed as follows.
\begin{align}
&\overline{\mathrm{BHAR}}(M_\star, z)=\nonumber\\
&\int_{\log\lambda_\mathrm{min}}^{+\infty}\frac{(1-\epsilon)k_\mathrm{bol}(M_\star\lambda)M_\star\lambda}{\epsilon c^2}p(\lambda\mid M_\star, z)d\log\lambda,\label{eq: master_bhar}
\end{align}
where $\epsilon$ is the radiative efficiency of the accretion. The key step in measuring $\overline{\mathrm{BHAR}}$ is hence to derive $p(\lambda\mid M_\star, z)$. Some literature models the $L_\mathrm{X}$ distribution instead of $\lambda$ (e.g., \citealt{Aird12}). These two approaches are equivalent, and $p(\lambda\mid M_\star, z)$ and $p(L_\mathrm{X}\mid M_\star, z)$ are interchangeable. The only reason for choosing one instead of the other is for convenience, as $\lambda$ is a scaled parameter that can serve as a rough proxy for the Eddington ratio.

\subsection{Semiparametric Modeling of $p(\lambda\mid M_\star, z)$}
\label{sec: plambda}
We assume a double power-law with respect to $\lambda$ for $p(\lambda\mid M_\star, z)$:
\begin{align}
p(\lambda\mid M_\star, z)=
\begin{cases}
A\left(\frac{\lambda}{\lambda_c}\right)^{-\gamma_1}, \lambda_\mathrm{min}<\lambda\leq\lambda_c\\
A\left(\frac{\lambda}{\lambda_c}\right)^{-\gamma_2}, \lambda>\lambda_c
\end{cases}\label{eq: plambda}
\end{align}
The four parameters $(A, \lambda_c, \gamma_1, \gamma_2)$ are functions of $(M_\star, z)$. We require $\lambda_c>\lambda_\mathrm{min}$ because, otherwise, the model will always degenerate to a single power law and has no dependence on $\gamma_1$ once $\lambda_c$ lies below $\lambda_\mathrm{min}$. We also require $\gamma_2>0$; otherwise, $p(\lambda\mid M_\star, z)$ will not be a probability measure, and the model-predicted number of AGNs will diverge.\footnote{Note that $p(\lambda\mid M_\star, z)$ is defined in the $\log\lambda$ space, and thus $\gamma_2>0$ is sufficient and necessary for $\int_{\log\lambda_\mathrm{min}}^{+\infty}p(\lambda\mid M_\star, z)d\log\lambda<+\infty$.} It has been shown that a double power-law can indeed approximate $p(\lambda\mid M_\star, z)$ well (e.g., \citealt{Bongiorno16, Aird18, Yang18a}). Similarly, the observed AGN \mbox{X-ray} luminosity function (XLF) also follows a double power-law (e.g., \citealt{Ueda14}), and a $p(\lambda\mid M_\star, z)$ roughly with a double power-law shape is needed to reproduce the XLF (Section~\ref{sec: hmcsample}).

\subsubsection{The Detection Probability}
\label{sec: Pdet}
We denote $P_\mathrm{det}(f_\mathrm{X})$ as the probability that a source with a $2-10$~keV flux of $f_\mathrm{X}$ is detected by a given \mbox{X-ray} survey. Following Section~3.4 in \citet{Zou23}, we adopt the following functional form for $P_\mathrm{det}(f_\mathrm{X})$:
\begin{align}
P_\mathrm{det}(f_\mathrm{X})=\frac{1}{2}\left[\mathrm{erf}\left(b(\log f_\mathrm{X}-a)\right)+1\right],\label{eq: Pdet}
\end{align}
where $a$ and $b$ are parameters determining the shape of $P_\mathrm{det}(f_\mathrm{X})$. We follow the same procedures as in \citet{Zou23} to calibrate $a$ and $b$ and report the results in Table~\ref{tbl_data}. Briefly, we compared the $f_\mathrm{X}$ distribution with the $2-10$~keV $\log N-\log S$ relation in \citet{Georgakakis08}, which is the well-determined expected surface number density per unit $f_\mathrm{X}$ with the detection procedures deconvolved. The comparison can return best-fit $(a, b)$ parameters such that the convolution between the $\log N-\log S$ relation and $P_\mathrm{det}$ best matches the observed $f_\mathrm{X}$ distribution. It is necessary to adopt a functional form because it improves the computational speed by several orders of magnitude, as will be discussed below. The form of Equation~\ref{eq: Pdet} has been shown to be appropriate for \mbox{X-ray} surveys (e.g., \citealt{Yan23, Zou23}) because its overall shape is similar to \mbox{X-ray} sensitivity curves, and, in our case, it indeed returns consistent $\overline{\mathrm{BHAR}}$ as in \citet{Yang18a}, who did not adopt this functional form for $P_\mathrm{det}$.\par
There is a subtle difference between eFEDS and the other fields. For the latter, their $f_\mathrm{X}$ is the observed value taken from the original \mbox{X-ray} catalogs. The $\log N-\log S$ relation is also for the observed $f_\mathrm{X}$; thus, $P_\mathrm{det}$ is for the observed $f_\mathrm{X}$. For eFEDS, we adopt the intrinsic, de-absorbed $0.5-2$~keV flux in \citet{Liu22} and multiply it by 1.57 to convert it to the \textit{intrinsic} $2-10$~keV flux assuming a photon-index of $\Gamma=1.8$. For consistency, we should correct the $\log N-\log S$ relation such that it works for the intrinsic $f_\mathrm{X}$. We use the XLF ($\phi_L$) in \citet{Ueda14} to derive the correction. The XLF-predicted intrinsic $\log N-\log S$ relation is
\begin{align}
N(f_\mathrm{X, int}>S)&=A_\mathrm{all~sky}^{-1}\int_{\log S}^{+\infty}\int_0^5\phi_L\left(L_\mathrm{X}, z\right)\frac{dV_C}{dz}d\log f_\mathrm{X, int}dz,\\
L_\mathrm{X}(f_\mathrm{X, int}, z)&=\frac{f_\mathrm{X, int}}{\eta(z)},\\
\eta(z)&=\frac{(1+z)^{2-\Gamma}}{4\pi D_L^2}\label{eq: eta}
\end{align}
where $A_\mathrm{all~sky}$ is the all-sky solid angle, $V_C$ is the comoving volume within a redshift of $z$, $\eta(z)$ is a function of $z$ converting $L_\mathrm{X}$ to the intrinsic $2-10$~keV flux for a power-law \mbox{X-ray} spectrum with a power-law photon index of $\Gamma=1.8$, and $D_L$ is the luminosity distance. We limit the integration to $z<5$ because the contribution of higher-redshift sources to the total source number is negligible. Similarly, the predicted observed $\log N-\log S$ relation is
\begin{align}
N(f_\mathrm{X, obs}>S)&=A_\mathrm{all~sky}^{-1}\int_{\log S}^{+\infty}\int_0^5\int_{20}^{24}\phi_L\left(L_\mathrm{X}, z\right)\frac{dV_C}{dz}\nonumber\\
&\times p\left(N_\mathrm{H}\mid L_\mathrm{X}, z\right)d\log f_\mathrm{X, obs}dzd\log N_\mathrm{H},\label{eq: logNlogSobs_pred}\\
L_\mathrm{X}(f_\mathrm{X, obs}, z, N_\mathrm{H})&=\frac{f_\mathrm{X, obs}}{\eta(z)\alpha(N_\mathrm{H}, z)},
\end{align}
where the $N_\mathrm{H}$ function $p\left(N_\mathrm{H}\mid L_\mathrm{X}, z\right)$ is the conditional probability density per unit $\log N_\mathrm{H}$ of an AGN with $(L_\mathrm{X}, z)$, as given in Section~3 of \citet{Ueda14}. This function is normalized such that $\int_{20}^{24}p\left(N_\mathrm{H}\mid L_\mathrm{X}, z\right)d\log N_\mathrm{H}=1$. $\alpha(N_\mathrm{H}, z)$ is the absorption factor for a source with $\Gamma=1.8$ and is calculated based on photoelectric absorption and Compton-scattering losses (i.e., \texttt{zphabs$\times$cabs}) in \texttt{XSPEC}.\par
The XLF-predicted $N(f_\mathrm{X, obs}>S)$ is similar to the observed $\log N-\log S$ relation, with an absolute difference generally below 0.2~dex. We found that $\log\left[N(f_\mathrm{X, int}>S)/N(f_\mathrm{X, obs}>S)\right]$ is almost a constant around 0.15~dex at $\log S>10^{-14}~\mathrm{erg~cm^{-2}~s^{-1}}$, and thus we add 0.15~dex to the observed $\log N-\log S$ relation in \citet{Georgakakis08} to approximate the intrinsic relation. Applying this intrinsic relation for our calibration in Equation~\ref{eq: Pdet}, we can obtain the eFEDS $P_\mathrm{det}$ as a function of the intrinsic $f_\mathrm{X}$. Given that the intrinsic $f_\mathrm{X}$ instead of the observed $f_\mathrm{X}$ is always adopted in our analyses of eFEDS, the fact that eFEDS is more easily affected by absorption has been appropriately accounted for and absorbed into $P_\mathrm{det}$. For example, the fact that obscured AGNs may be missed by eFEDS causes the $a$ value to slightly shift to a larger value due to the correction applied to the observed $\log N-\log S$ relation. One may wonder why we convert the $0.5-2$~keV flux to $2-10$~keV flux instead of directly using $0.5-2$~keV flux. Since the intrinsic flux is always adopted, the conversion, in principle, would not cause systematic biases. The main reason is that the correction to the $\log N-\log S$ relation is considerably smaller for the $2-10$~keV band than for the $0.5-2$~keV band.\par
One caveat is that we limit the integration range of $N_\mathrm{H}$ in Equation~\ref{eq: logNlogSobs_pred} to be below $10^{24}~\mathrm{cm^{-2}}$, which equivalently means that we neglect the contribution from Compton-thick (CT) AGNs with $N_\mathrm{H}>10^{24}~\mathrm{cm^{-2}}$ for eFEDS. Similarly, in our other fields observed by Chandra or XMM-Newton, we also implicitly neglect most CT AGNs because they can hardly be detected even in the HB. Generally, \mbox{X-ray} observations below 10~keV cannot provide effective constraints for the CT population, and the intrinsic fraction of CT AGNs is highly uncertain (e.g., \citealt{Ananna19}). Therefore, any attempt to measure the intrinsic CT population properties using \mbox{X-rays} below 10~keV is likely prone to large systematic uncertainties. The CT population might indeed contribute to the SMBH growth and is missed by our measurements, especially at high redshift (e.g., \citealt{Yang21}), but observations in the regime insensitive to the CT obscuration are necessary to reveal it (e.g., \citealt{Yang23}).

\subsubsection{The Likelihood}
When compared with the observed data, the log-likelihood function (e.g., \citealt{Loredo04}) is
\begin{align}
\ln\mathcal{L}&=-\sum_{s=1}^{n^\mathrm{gal}}T_{\mathrm{gal}, s}+\sum_{s=1}^{n^\mathrm{AGN}}\ln p(\lambda_s\mid M_{\star,s}, z_s),\label{eq: lnL}\\
T_\mathrm{gal}&=\int_{\log\lambda_\mathrm{min}}^{+\infty}p(\lambda\mid M_\star, z)P_\mathrm{det}(f_\mathrm{X}(\lambda M_\star, z))d\log\lambda,\label{eq: Tgal}\\
f_\mathrm{X}&(\lambda M_\star, z)=\lambda M_\star\eta(z),
\end{align}
where $\eta(z)$ is given in Equation~\ref{eq: eta}. We adopt $\Gamma=1.8$ and 1.6 for eFEDS and the other fields, respectively. Different $\Gamma$ values are adopted because the adopted $f_\mathrm{X}$ inside our $P_\mathrm{det}$ function is the intrinsic value for eFEDS while being the observed one for the other fields (Section~\ref{sec: Pdet}). Equation~\ref{eq: Tgal} involves an integration, and Equation~\ref{eq: lnL} computes Equation~\ref{eq: Tgal} many times in the summation for a single evaluation of $\mathcal{L}$. Numerically integrating Equation~\ref{eq: Tgal} is slow, making it impractical to sample more than one or two dozen free parameters (as will be shown later, we will have $10^4$ free parameters). Fortunately, as previously suggested in \citet{Zou23}, Equation~\ref{eq: Tgal} can be analytically solved when choosing appropriate functional forms for $p(\lambda\mid M_\star, z)$ and $P_\mathrm{det}(f_\mathrm{X})$, and our Equations~\ref{eq: plambda} and \ref{eq: Pdet} enable this. This is one of the most important steps enabling our whole semiparametric analyses.\par
We define
\begin{align}
&I(\gamma, \lambda_1, \lambda_2, A, \lambda_c; M_\star, z)=\nonumber\\
&\int_{\log\lambda_1}^{\log\lambda_2}A\left(\frac{\lambda}{\lambda_c}\right)^{-\gamma}P_\mathrm{det}(\lambda M_\star\eta(z))d\log\lambda,\label{eq: I}
\end{align}
Using Equation~21 in \citet{Zou23}, Equation~\ref{eq: I} can be reduced as follows.
\begin{align}
I&=\frac{A}{2\gamma\ln10}\Biggl\{\left(\frac{\lambda_1}{\lambda_c}\right)^{-\gamma}\left[\mathrm{erf}(x_1)+1\right]-\left(\frac{\lambda_2}{\lambda_c}\right)^{-\gamma}\left[\mathrm{erf}(x_2)+1\right]\nonumber\\
&-\left(\frac{10^{a-\frac{\gamma\ln10}{4b^2}}}{\lambda_cM_\star\eta}\right)^{-\gamma}\left[\mathrm{erf}\left(x_1+\frac{\gamma\ln10}{2b}\right)-\mathrm{erf}\left(x_2+\frac{\gamma\ln10}{2b}\right)\right]\Biggl\},\\
x_k&=b\left[\log(\lambda_kM_\star\eta)-a\right], k=1, 2.
\end{align}
Equation~\ref{eq: Tgal} can then be expressed as follows.
\begin{align}
T_\mathrm{gal}(A, \lambda_c, \gamma_1, \gamma_2; M_\star, z)&=I(\gamma_1, \lambda_\mathrm{min}, \lambda_c, A, \lambda_c; M_\star, z)\nonumber\\
&+I(\gamma_2, \lambda_c, +\infty, A, \lambda_c; M_\star, z).
\end{align}\par
The above equations express $\mathcal{L}$ as a function of $(A, \lambda_c, \gamma_1, \gamma_2)$, which themselves are functions of $(M_\star, z)$. The dependences of $(A, \lambda_c, \gamma_1, \gamma_2)$ on $(M_\star, z)$ lack clear guidelines, and we use a nonparametric approach to model them. We divide the $(M_\star, z)$ plane into $N_M\times N_z$ grids and adopt the $(A, \lambda_c, \gamma_1, \gamma_2)$ values in each grid element as free parameters; i.e., we have $4N_MN_z$ free parameters in total. Such an approach is conceptually similar to and was indeed initially inspired by the ``gold standard'' nonparametric star-formation history (e.g., \citealt{Leja19}) in SED fitting. In a strict statistical sense, a method is called ``nonparametric'' only if the number of free parameters scales with the number of data points. In contrast, we used a fixed number of free parameters, which does not exactly satisfy the statistical definition. Although we can easily adjust $N_M$ and $N_z$ so that the number of free parameters scales with the number of data points, this makes the computation infeasible because we have millions of galaxies; besides, with our continuity prior in Section~\ref{sec: prior}, further increasing the number of free parameters does not improve our results materially. In our context, we use the word ``nonparametric'' because our number of free parameters is far larger than that of typical parametric methods, and our method is effectively similar to the fully nonparametric approach. This same argument also works for ``nonparametric star-formation history'' in, e.g., \citet{Leja19}.\par
This method has an important advantage over a parametric one in our case. As Figure~\ref{fig_dataplot_total} shows, most of our data are clustered within a small region of the $(M_\star, z)$ plane -- the number of sources significantly decreases at both low $z$ ($\lesssim0.8$) and high $z$ ($\gtrsim2$), the number of galaxies strongly depends on $M_\star$, and most AGNs are confined within $10^{10.5}\lesssim M_\star\lesssim10^{11.2}~M_\odot$. This indicates that if we assume any parametric form for $(A, \lambda_c, \gamma_1, \gamma_2)$, the fitted parameters will be dominated by the small but well-populated region in the $(M_\star, z)$ plane. Especially, one strong argument disfavoring parametric fitting is that our ultimate goal is to derive $\overline{\mathrm{BHAR}}$ across all redshifts, but any parametric fitting will return results dominated by sources in a small redshift range (e.g., \citealt{Yang18a}). Our semiparametric settings avoid this problem.\par
Equation~\ref{eq: lnL} then becomes
\begin{align}
\ln\mathcal{L}=\sum_{i=1}^{N_M}\sum_{j=1}^{N_z}&\left[-n_{ij}^\mathrm{gal}T_\mathrm{gal}(A_{ij}, \lambda_{c, ij}, \gamma_{1, ij}, \gamma_{2, ij}; M_{\star, i}, z_j)\right.\nonumber\\
&+\sum_{s=1}^{n_{ij}^\mathrm{AGN}}\left.\ln p_{ij}(\lambda_s\mid M_{\star, s}, z_s)\right],\label{eq: lnlike_final}
\end{align}
where $n_{ij}^\mathrm{gal}$ and $n_{ij}^\mathrm{AGN}$ are the numbers of galaxies and AGNs within the $(i, j)$ bin, respectively. $\mathcal{L}$ is defined for each individual survey field, and they are added together to return the final likelihood.\par

\subsubsection{The Prior}
\label{sec: prior}
We adopt a continuity prior:
\begin{align}
X_{i+1,j}-X_{ij}&\sim N\left(0, \frac{\sigma_X^2}{N_M}\right),\\
X_{i,j+1}-X_{ij}&\sim N\left(0, \frac{\sigma_X^2}{N_z}\right),
\end{align}
where $X$ denotes each one of $(\log A, \log\lambda_c, \gamma_1, \gamma_2)$, and $\sigma_X$ is our chosen a priori parameters to quantify the overall variations of $X$ across the whole fitting ranges. The goal of this continuity prior is to transport information among grid elements. Without this prior, the fitted parameters in each grid element become unstable and vary strongly. This prior is defined in a way such that the information flow is roughly independent of the grid size. The continuity prior is defined only for the differences, and we need a further prior for the $X$'s in a single cell and adopt it as flat in the $(\log A, \log\lambda_c, \gamma_1, \gamma_2)$ space. We set bounds for these parameters to ensure propriety of the prior \citep{Tak18}: $-10<\log A<10$, $\log\lambda_\mathrm{min}<\log\lambda_c<40$, $-5<\gamma_1<10$, and $0<\gamma_2<10$. These ranges are sufficiently large to encompass any reasonable parameter values. Our posterior (Section~\ref{sec: post}) may also become less numerically stable outside these bounds. The resulting prior is explicitly shown below.
\begin{align}
\ln\pi_\mathrm{cont}=-\frac{1}{2}\sum_X&\left[N_M\sum_{i=1}^{N_M-1}\sum_{j=1}^{N_z}\frac{(X_{i+1,j}-X_{ij})^2}{\sigma_X^2}\right.\nonumber\\
&\left.+N_z\sum_{i=1}^{N_M}\sum_{j=1}^{N_z-1}\frac{(X_{i,j+1}-X_{ij})^2}{\sigma_X^2}\right].\label{eq: lnprior}
\end{align}
Note that it is defined in the $(\log A, \log\lambda_c, \gamma_1, \gamma_2)$ space, and an appropriate Jacobian determinant should be added when transforming the parameter space. For sampling purposes, variable transformations are usually needed.\par
We rely on previous literature to set appropriate values for $\sigma_X$. \citet{Yang18a} used a double power-law similar to ours to fit $p(\lambda\mid M_\star, z)$, and their best-fit parameters (see their Equation~16) span ranges of $-3.53<\log A<-0.86, 31.73<\log\lambda_c<34.98, \gamma_1=0.43, 1.55<\gamma_2<3.55$ across our parameter spaces. \citet{Bongiorno16} modeled the bivariate distribution function of $M_\star$ and $\lambda$ for AGNs, which can be converted to $p(\lambda\mid M_\star, z)$ by dividing it by the galaxy stellar mass function (SMF), and the corresponding $p(\lambda\mid M_\star, z)$ is also a double power-law. We use the SMF in \citet{Wright18} for the conversion, and the best-fit double power-law parameters in \citet{Bongiorno16} span ranges of $-5.28<\log A<-0.08, 33.32<\log\lambda_c<34.52, -0.67<\gamma_1<1.62, \gamma_2=3.72$. \citet{Aird18} nonparametrically modeled $p(\lambda\mid M_\star, z)$, and we use our double power-law model to fit their results above $M_\star=10^{9.5}~M_\odot$ by minimizing the Kullback-Leibler divergence of our model relative to theirs. The returned best-fit values range between $-2.87<\log A<-0.69, 31.84<\log\lambda_c<34.04, -0.58<\gamma_1<0.52, 0.72<\gamma_2<1.67$. Another independent way to estimate $p(\lambda\mid M_\star, z)$ is based on the fact that $p(\lambda\mid M_\star, z)$, by definition, can predict the XLF when combined with the SMF (see Equation~\ref{eq: smf_to_xlf} and Section~\ref{sec: addxlf} for more details). We estimate parameters of $p(\lambda\mid M_\star, z)$ such that, when using the SMF in \citet{Wright18}, the predicted XLF can match the best with the XLF in \citet{Ueda14}. This returns $-2.81<\log A<-1.08, 32.72<\log\lambda_c<33.77, -0.35<\gamma_1<0.90, 2.46<\gamma_2<2.82$. Taking the union of these estimations, the ranges should span no more than $-5.28<\log A<-0.08, 31.73<\log\lambda_c<34.98, -0.67<\gamma_1<1.62, 0.72<\gamma_2<3.72$. We adopt $\sigma_X$ as one third of the widths,\footnote{A nominal $\sigma$ is often approximated by $1/4$ of the range, according to the so-called \textit{range rule of thumb}. We have two dimensions in our case, and thus the one-dimension $\sigma$ can be chosen as $1/\left(4\sqrt{2}\right)$ of the range. However, we would like to be slightly more conservative. The reason is that previous works mostly do not cover a parameter space as large as this work, and thus extrapolations are employed when computing the ranges. Some conservativeness can enable more flexibility to accommodate possible systematic extrapolation errors in regimes not well-covered by previous works.} i.e., $\sigma_{\log A}=1.7, \sigma_{\log\lambda_c}=1.1, \sigma_{\gamma_1}=0.8, \sigma_{\gamma_2}=1.0$.\par
In fact, our prior setting is essentially a rasterized approximation to the continuous surface of a Gaussian process (GP) regression (e.g., \citealt{Rasmussen06}). This is because the blocky prior surface over the $(M_\star, z)$ plane becomes the non-parametric GP-based continuous surface as the resolution of the grid increases (i.e., increasing $N_M$ and $N_z$ to the infinity). Therefore, a full GP regression involves a large number of free parameters scaling with the galaxy sample size ($\approx10^6$), while our rasterized approach only involves $10^4$ parameters. GP also requires computations of $\mathcal{O}(n^3)$ for matrix inversions, while our approach turns the matrix-inversion problem into products of multiple univariate Gaussian densities. Due to these reasons, a full GP regression is computationally infeasible in our case, but our approach effectively works similarly and is much less computationally demanding.

\subsubsection{The Posterior}
\label{sec: post}
The posterior is
\begin{align}
\ln\mathcal{P}=\sum_\mathrm{field}\ln\mathcal{L}+\ln\pi_\mathrm{cont}.\label{eq: lnpost}
\end{align}
We call our overall modeling ``semiparametric'' because we adopt $p(\lambda\mid M_\star, z)$ as a parametric function of $\lambda$, while the dependences of $(A, \lambda_c, \gamma_1, \gamma_2)$ on $(M_\star, z)$ are nonparametric. Readers may wonder why we do not also adopt a nonparametric function for $p(\lambda\mid M_\star, z)$. In principle, it could be done and was presented in \citet{Georgakakis17} and \citet{Aird18}. Since any model contains subjective assumptions, the choice of the methodology should be guided by the assumptions we want to retain or avoid. Compared to nonparametric modeling, the assumptions of parametric models are much stronger. We nonparametrically model $(A, \lambda_c, \gamma_1, \gamma_2)$ as functions of $(M_\star, z)$ because we genuinely do not know their dependencies and thus want to minimize assumptions. However, we are satisfied with and thus want to retain the inherent assumption of our parametrization of $p(\lambda\mid M_\star, z)$ that the true dependence is indeed well-approximated by a double power-law when $\lambda>\lambda_\mathrm{min}$. Previous works have shown that a double power-law indeed works, and, as far as we know, there is no clear evidence suggesting that this assumption would fail. Especially, the nonparametric form of $p(\lambda\mid M_\star, z)$ inferred from \citet{Aird18} is also roughly a double power-law. The adopted approach essentially depends on our ultimate goal. It is certainly better to minimize the assumption for $p(\lambda\mid M_\star, z)$ and adopt a nonparametric form for it if the ultimate goal is to derive the shape of $p(\lambda\mid M_\star, z)$. However, our goal is different -- we are ultimately interested in $\overline{\mathrm{BHAR}}$ and thus want to assume a double power-law form for $p(\lambda\mid M_\star, z)$.

\subsection{Hamiltonian Monte Carlo Sampling of $p(\lambda\mid M_\star, z)$}
\label{sec: hmcsample}
\begin{figure*}
\centering
\resizebox{\hsize}{!}{\includegraphics{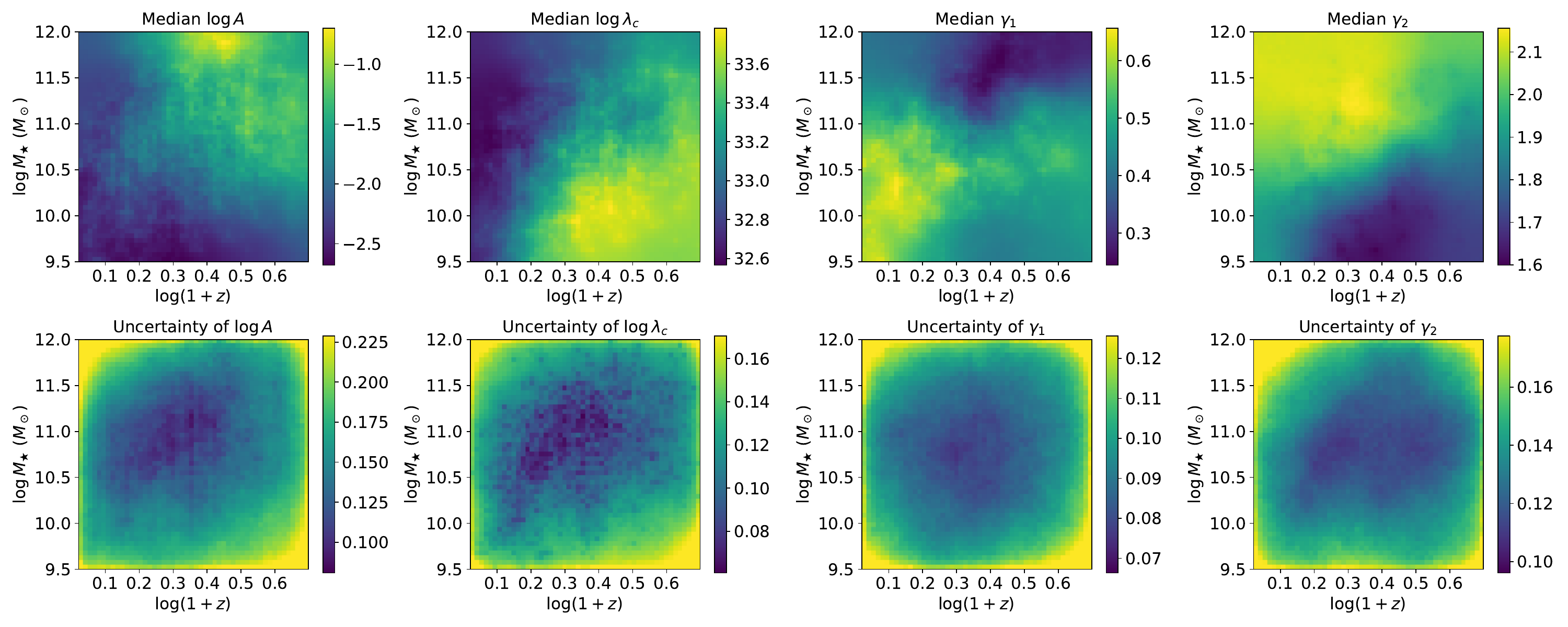}}
\caption{The sampled maps of $(A, \lambda_c, \gamma_1, \gamma_2)$. The top panels are the median posteriors, and the bottom panels are the $1\sigma$ uncertainties, defined as the half width of the posterior's 16th-84th percentile range.}
\label{fig_thetamap}
\end{figure*}

Given the high dimensionality, Hamiltonian Monte Carlo (HMC; e.g., \citealt{Betancourt17}) should be one of the most practical methods to sample the posterior. As far as we know, other sampling methods either cannot work efficiently in our high-dimension case (e.g., the traditional Metropolis–Hastings algorithm) or do not have well-developed packages readily available (e.g., \citealt{Bayer23}). HMC needs both the posterior and its gradient in the parameter space. The posterior has been presented in the previous subsections, and we present the gradient in Appendix~\ref{append: gradient}. We use \texttt{DynamicHMC.jl}\footnote{\url{https://www.tamaspapp.eu/DynamicHMC.jl/stable/}} to conduct the HMC sampling. We adopt $N_M=49$ and $N_z=50$. The sampling results are presented in Figure~\ref{fig_thetamap}. These parameter maps will be released online.\par

\begin{figure*}
\centering
\resizebox{\hsize}{!}{\includegraphics{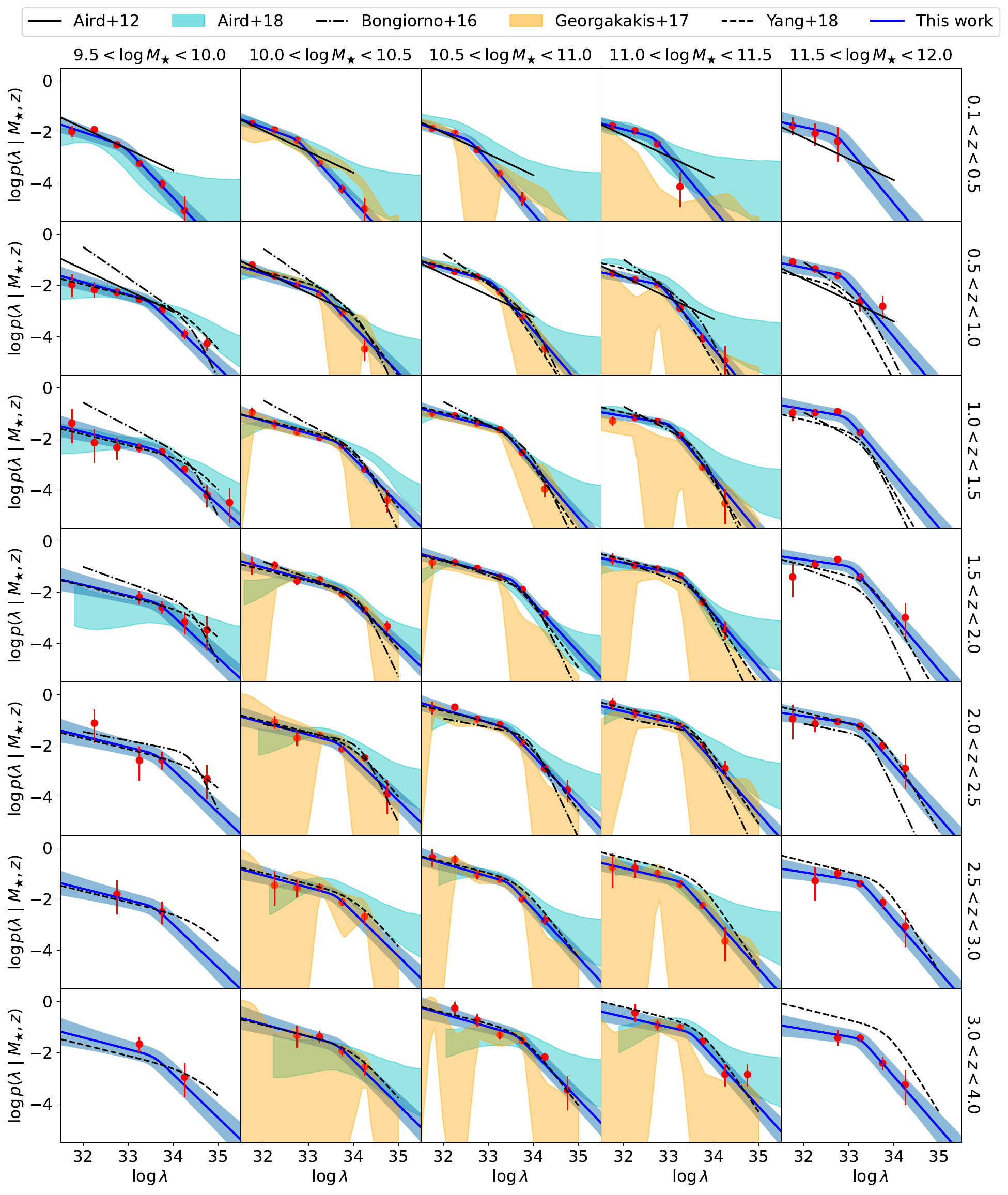}}
\caption{Comparison between our $p(\lambda\mid M_\star, z)$ and other measurements. The red points are the binned estimators with $1\sigma$ error bars based on our data. The blue curves are our fitted median $p(\lambda\mid M_\star, z)$, evaluated at the bin centers, and the blue shaded regions are the corresponding 90\% confidence ranges. The black solid straight lines are the single power-law models in \citet{Aird12}. The dash-dotted and dashed curves are the double power-law models in \citet{Bongiorno16} and \citet{Yang18a}, respectively. The cyan and orange shaded regions are the 90\% confidence intervals of the nonparametric $p(\lambda\mid M_\star, z)$ in \citet{Aird18} and \citet{Georgakakis17}, respectively.}
\label{fig_plambda_literature}
\end{figure*}

To examine our fitting quality, we compare the model $p(\lambda\mid M_\star, z)$ with the observed values. We use the $n_\mathrm{obs}/n_\mathrm{mdl}$ method to obtain binned estimators of $p(\lambda\mid M_\star, z)$, as outlined in \citet{Aird12}. For a given $(z, M_\star, \lambda)$ bin ranging $[z_\mathrm{low}, z_\mathrm{high}]\times[M_{\star, \mathrm{low}}, M_{\star, \mathrm{high}}]\times[\lambda_\mathrm{low}, \lambda_\mathrm{high}]$, we denote the number of observed AGNs as $n_\mathrm{obs}$ and calculate the model-predicted number as $n_\mathrm{mdl}$:
\begin{align}
n_\mathrm{mdl}=\sum_s\int_{\log\lambda_\mathrm{low}}^{\log\lambda_\mathrm{high}}p(\lambda\mid M_{\star, s}, z_s)P_\mathrm{det}(f_\mathrm{X}(\lambda M_{\star, s}, z_s))d\log\lambda,
\end{align}
where the summation runs over all the sources within $[z_\mathrm{low}, z_\mathrm{high}]\times[M_{\star, \mathrm{low}}, M_{\star, \mathrm{high}}]$. The observed binned estimator of $p(\lambda\mid M_\star, z)$ is then the fitted model evaluated at the bin center scaled by $n_\mathrm{obs}/n_\mathrm{mdl}$, and its uncertainty is calculated from the Poisson error of $n_\mathrm{obs}$ following \citet{Gehrels86}. We present our model $p(\lambda\mid M_\star, z)$ and the binned estimators in Figure~\ref{fig_plambda_literature}, and they are consistent. The uncertainties become large especially in the high-$z$/low-$M_\star$ and low-$z$/high-$M_\star$ regimes because of a limited number of AGNs being available. In the high-$z$/low-$M_\star$ regime, most of the constraints are from deep CANDELS fields, especially GOODS-S, because the other fields are not sufficiently deep in both \mbox{X-rays} and other multi-wavelength bands. For example, 60\% (80\%) of AGNs in our sample with $M_\star<10^{10}~M_\odot$ and $z>2$ ($z>3$) are from GOODS-S. At $z<0.5$, $\gtrsim60\%$ of AGNs are from eFEDS, and even the $60~\mathrm{deg^2}$ eFEDS is not sufficiently large to effectively sample high-$M_\star$ sources at low redshift. We also plot several $p(\lambda\mid M_\star, z)$ results from previous works and leave more detailed discussions on the comparison between our $p(\lambda\mid M_\star, z)$ and previous ones to Section~\ref{sec: compare_literature}.\par

\begin{figure*}
\centering
\resizebox{\hsize}{!}{\includegraphics{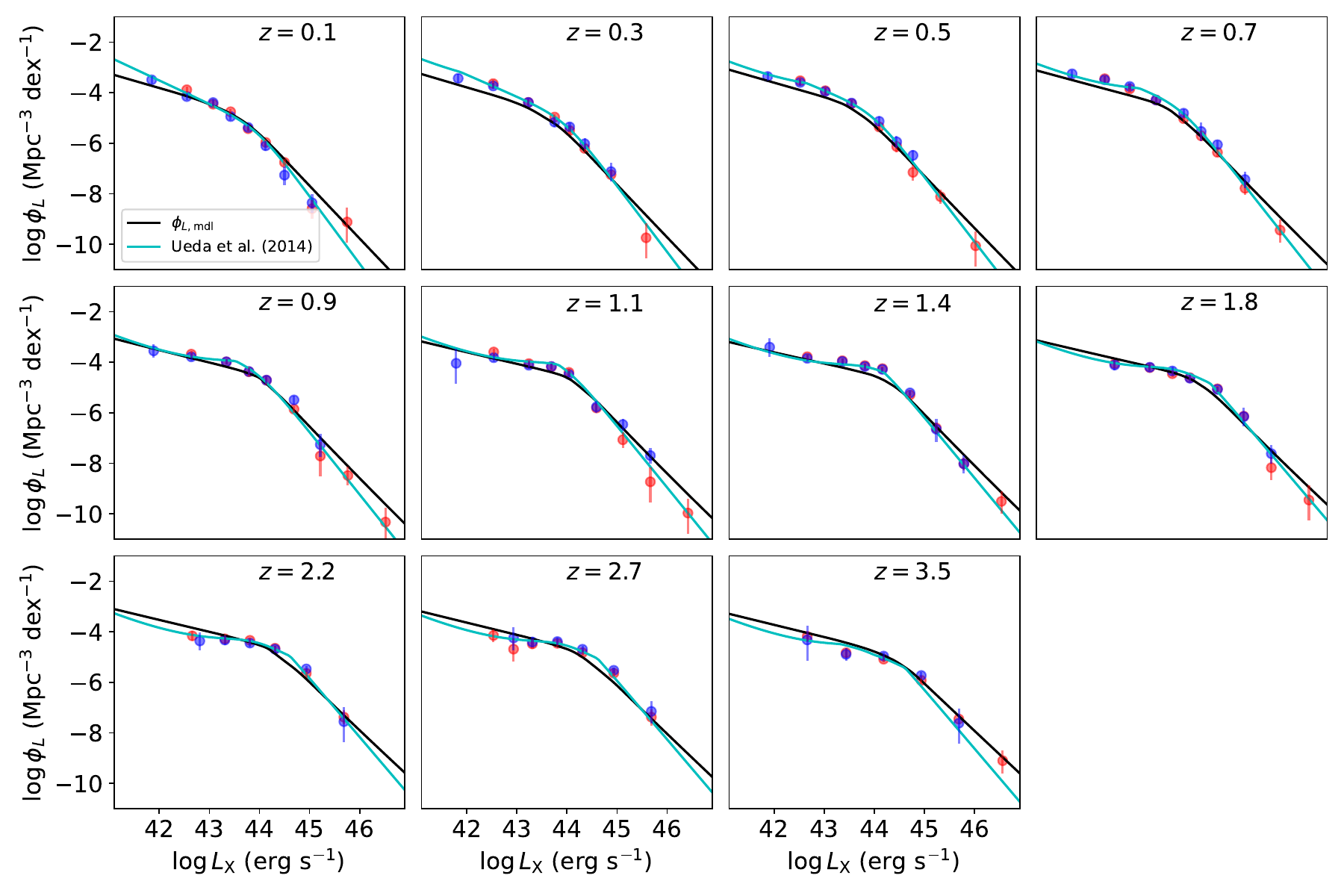}}
\caption{The XLFs at different redshifts. The red (blue) data points are the soft-band (hard-band) XLFs in \citet{Ueda14}. The cyan curves are the best-fit XLF models in \citet{Ueda14}, and the black curves are our $\phi_{L, \mathrm{mdl}}$ based on the median parameter maps in Figure~\ref{fig_thetamap} and the SMF in \citet{Wright18}. The absorption correction has been appropriately applied for both our measurements (see Section~\ref{sec: Pdet}) and the XLFs in \citet{Ueda14}. Our models agree with the observed XLFs well.}
\label{fig_xlf}
\end{figure*}

As another independent check, $p(\lambda\mid M_\star, z)$, by definition, can connect the SMF ($\phi_M$) and XLF ($\phi_L$). That is, the SMF and $p(\lambda\mid M_\star, z)$ can jointly predict the XLF (e.g., \citealt{Bongiorno16, Georgakakis17}):
\begin{align}
\phi_{L, \mathrm{mdl}}(L_\mathrm{X}, z)&=\int_{\log M_{\star, \mathrm{min}}}^{\log M_{\star, \mathrm{max}}}p(\lambda\mid M_\star, z)\phi_Md\log M_\star\nonumber\\
&=\int_{\log M_{\star, \mathrm{min}}}^{\log M_{\star, \mathrm{max}}}p(L_\mathrm{X}/M_\star\mid M_\star, z)\phi_Md\log M_\star.\label{eq: smf_to_xlf}
\end{align}
Comparing $\phi_{L, \mathrm{mdl}}$ and the observed XLF, $\phi_{L, \mathrm{obs}}$, can thus further assess our fitting quality. We adopt $\phi_M$ in \citet{Wright18} and the median parameter maps in Figure~\ref{fig_thetamap} to calculate $\phi_{L, \mathrm{mdl}}$. We present the comparison between $\phi_{L, \mathrm{mdl}}$ and $\phi_{L, \mathrm{obs}}$ from \citet{Ueda14} in Figure~\ref{fig_xlf}, and they agree well. Note that for comparison purposes here, we do not need to optimize the computation of Equation~\ref{eq: smf_to_xlf}; however, we will present a more optimized computation algorithm later in Section~\ref{sec: addxlf}, where we do need fast computational speed. Also, note that Equation~\ref{eq: smf_to_xlf} ignores the contribution from sources with $M_\star$ below $M_{\star, \mathrm{min}}=10^{9.5}~M_\odot$ or above $M_{\star, \mathrm{max}}=10^{12}~M_\odot$ to the XLF. This is appropriate because the XLF is dominated by AGNs with $10^{9.5}<M_\star<10^{12}~M_\odot$. As a simple check, for the parameters in Figure~\ref{fig_thetamap}, if we extrapolate the integration in Equation~\ref{eq: smf_to_xlf} to $(-\infty, +\infty)$, the typical $\phi_{L, \mathrm{mdl}}$ will only increase by 0.01~dex at $43<L_\mathrm{X}<43.5$, the lowest $L_\mathrm{X}$ bin that we will later adopt in Section~\ref{sec: addxlf}. This increment is even smaller for higher $L_\mathrm{X}$ bins.\par

\subsection{Measuring $\overline{\mathrm{BHAR}}$}
\label{sec: bhar}
\begin{figure}
\centering
\resizebox{\hsize}{!}{\includegraphics{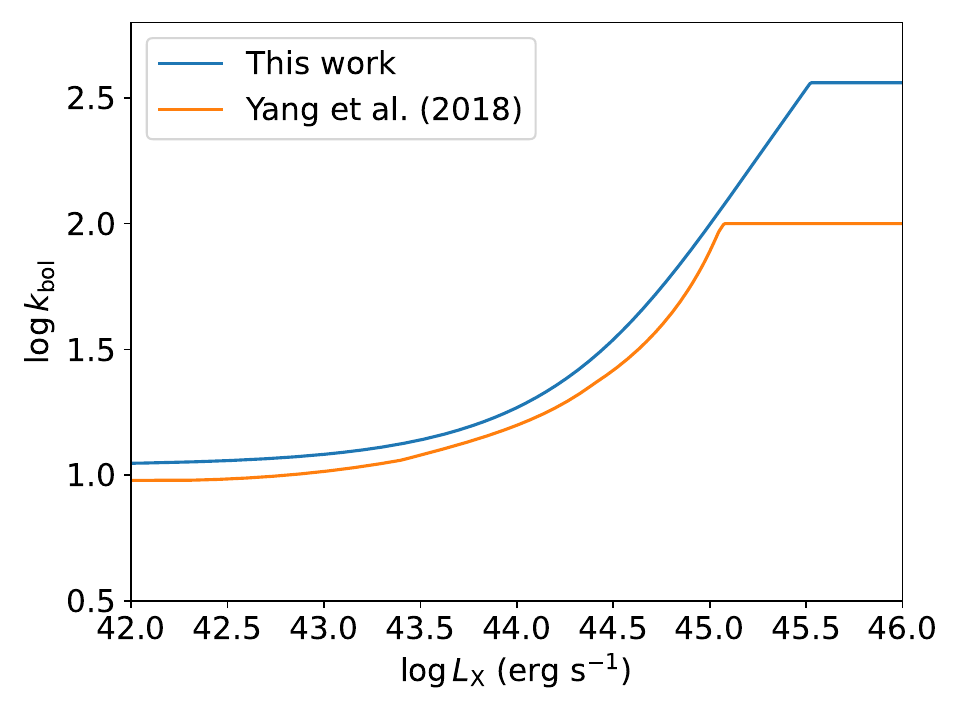}}
\caption{The adopted $L_\mathrm{X}-k_\mathrm{bol}$ relation, taken from \citet{Duras20}. The adopted relation used in \citet{Yang18a}, which is adjusted from \citet{Lusso12}, is also plotted for comparison.}
\label{fig_kbol}
\end{figure}

\begin{figure}
\centering
\resizebox{\hsize}{!}{\includegraphics{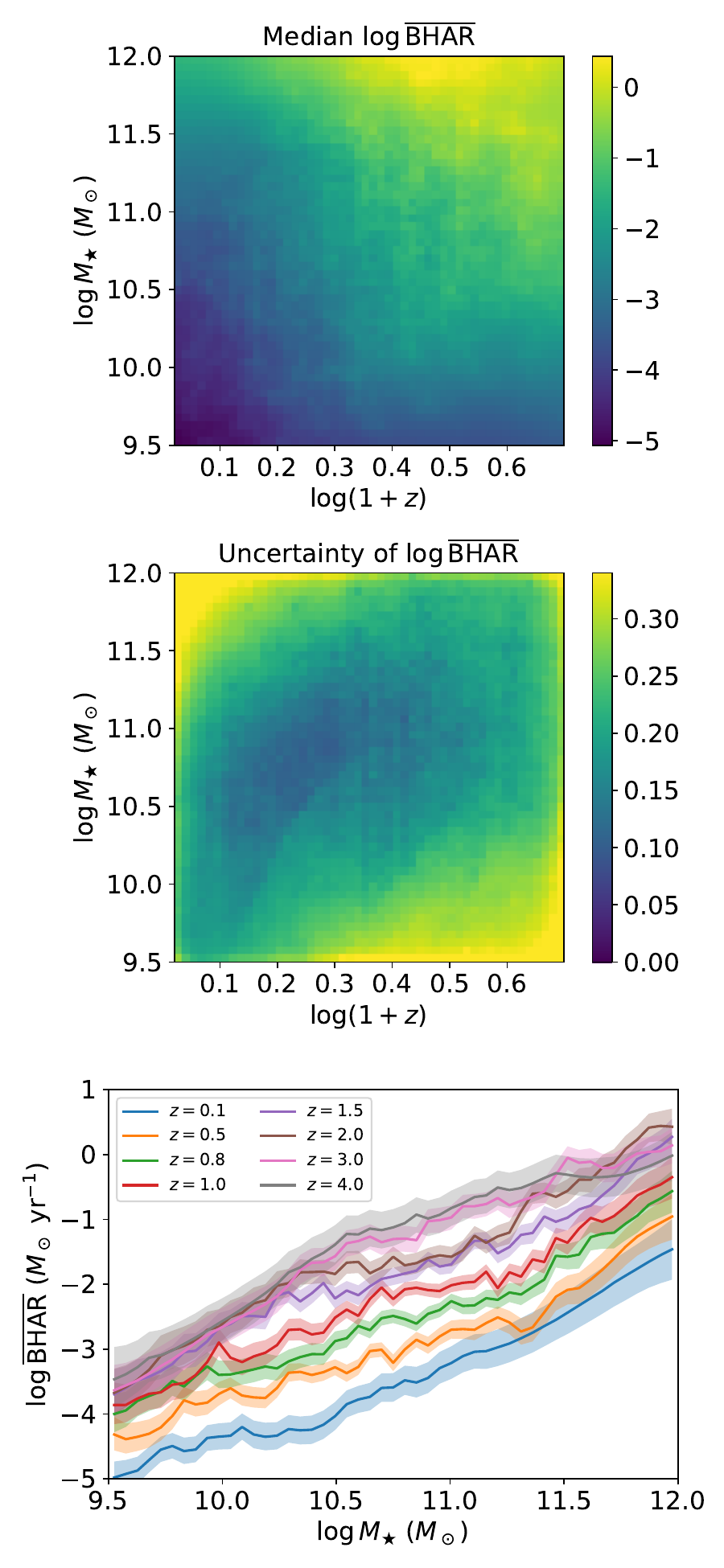}}
\caption{The top and middle panels are the sampled $\overline{\mathrm{BHAR}}$ maps, where the unit of $\overline{\mathrm{BHAR}}$ is $M_\odot~\mathrm{yr^{-1}}$. The bottom panel shows the $\overline{\mathrm{BHAR}}-M_\star$ relation at several redshifts, where the solid curves are the median values, and the shaded regions are the corresponding $1\sigma$ uncertainty ranges. $\overline{\mathrm{BHAR}}$ generally increases with both $M_\star$ and $z$.}
\label{fig_bharresult}
\end{figure}

Equation~\ref{eq: master_bhar} converts $p(\lambda\mid M_\star, z)$ to $\overline{\mathrm{BHAR}}$. We adopt $\epsilon=0.1$ and $k_\mathrm{bol}$ from Equation~2 in \citet{Duras20}. In principle, $\epsilon$ may depend upon other factors such as the accretion state (e.g., \citealt{Yuan14}), but it is infeasible to accurately measure $\epsilon$ for our individual sources. We adopt $\epsilon$ as 0.1 because it is a typical value for the general AGN population (e.g., \citealt{Brandt15}) and has been widely used in previous literature (e.g., \citealt{Yang17, Yang18a, Yang19, Ni19, Ni21b}). The $k_\mathrm{bol}$ relation in \citet{Duras20} diverges at high $L_\mathrm{X}$. To avoid it, we cap $k_\mathrm{bol}$ not to exceed 363, the value when the bolometric luminosity is $10^{14.5}~L_\odot$, which is roughly the brightest sample used in \citet{Duras20} to calibrate the $k_\mathrm{bol}$ relation. We show the $L_\mathrm{X}-k_\mathrm{bol}$ relation in Figure~\ref{fig_kbol}, in which we also plot the relation used in \citet{Yang18a}, derived from \citet{Lusso12}, for a comparison. The two relations are similar, with a small offset of $\approx0.07$~dex that is almost negligible compared to the $\overline{\mathrm{BHAR}}$ uncertainty (Figure~\ref{fig_bharresult}). The deviation of the two relations at $L_\mathrm{X}\gtrsim10^{45}~\mathrm{erg~s^{-1}}$ has little impact on $\overline{\mathrm{BHAR}}$ because $\overline{\mathrm{BHAR}}$ has little contribution from $\log\lambda\gtrsim35$ (see Figure~\ref{fig_plambda_literature}).\par

Equation~\ref{eq: master_bhar} ignores the contribution to $\overline{\mathrm{BHAR}}$ from sources at $\log\lambda<31.5$ because \mbox{X-ray} binaries may not be negligible at lower $\lambda$, and our \mbox{X-ray} surveys can hardly provide strong constraints in the low-$\lambda$ regime. However, this will not cause material biases because $\overline{\mathrm{BHAR}}$ is dominated by sources at $\log\lambda\gtrsim31.5$ (e.g., Section~3.2.3 in \citealt{Yang18a}). We have also tried pushing the lower integration bound in Equation~\ref{eq: master_bhar} down by 2~dex, and the returned $\overline{\mathrm{BHAR}}$ would only increase by a typical value of $\approx0.02$~dex and no more than $\approx0.1$~dex. Such a difference is much smaller than the fitted $\overline{\mathrm{BHAR}}$ uncertainty. This exercise may even overestimate the influence because $p(\lambda\mid M_\star, z)$ may bend downward or quickly vanish at very small $\lambda$ \citep{Aird17, Aird18}. Therefore, the cut at $\log\lambda=31.5$ is not expected to cause material biases to $\overline{\mathrm{BHAR}}$.\par
We show our sampled $\overline{\mathrm{BHAR}}$ results in Figure~\ref{fig_bharresult}, and the $\overline{\mathrm{BHAR}}$ maps will be released online. The median map clearly shows that $\overline{\mathrm{BHAR}}$ increases with both $M_\star$ and $z$, qualitatively consistent with the conclusions in \citet{Yang18a}. The uncertainty map reveals that the $\overline{\mathrm{BHAR}}$ constraints at both the low-$z$/high-$M_\star$ and the high-$z$/low-$M_\star$ regime are relatively more limited. We will present more quantitative comparison with \citet{Yang18a} and other works in Section~\ref{sec: compare_literature}. Besides, we verified that AGN-dominated sources do not cause material biases to our $\overline{\mathrm{BHAR}}$ measurements in Appendix~\ref{append: blagn}.\par

There are slight, local fluctuations in $\overline{\mathrm{BHAR}}$ that are caused by the statistical noise of the data and are confined within the extent allowed by our prior, and the $\overline{\mathrm{BHAR}}$ map is smooth globally, as can be seen in the top panel of Figure~\ref{fig_bharresult}. The fluctuation levels and $\overline{\mathrm{BHAR}}$ uncertainties depend on our prior settings but almost not on our bin size because our bins are set to be correlated (Section~\ref{sec: prior}). For example, relaxing the prior by choosing larger $\sigma_X$ would return larger fluctuations and uncertainties. This ``arbitrariness'' is inherent in modeling.\footnote{For the widely used method of binning the parameter space and assuming each bin is independent, there is a similar arbitrariness in choosing the bin size, and the uncertainties in this case would depend on the bin size.} Overall, our prior is reasonably constructed (Section~\ref{sec: prior}) and provides beneficial regularizations. We have assessed the potential issue of whether such arbitrary choices affect the following posterior inferences and the resulting scientific conclusions qualitatively. For example, we have conducted a sensitivity check of our priors and confirmed that the impact of lower or higher resolution of the prior surface (corresponding to larger or smaller bin sizes) does not influence the resulting posterior inference in a noticeable way, and changing $\sigma_X$ generally would not cause material changes of the median $\overline{\mathrm{BHAR}}$ map.

\section{Discussion}
\label{sec: discussion}
Given that this article is already lengthy and full of technical details, we decide to present more scientific investigations of our results in future dedicated works. However, we would like to present brief, immediate, but sufficiently informative explorations of our results in this section, which helps justify the quality and serves as a precursor of further detailed scientific investigations.

\subsection{Adding External Constraints from the SMF and XLF}
\label{sec: addxlf}
Section~\ref{sec: hmcsample} uses the SMF and XLF to examine the fitting quality of $p(\lambda\mid M_\star, z)$. It is also possible to follow a reversed direction -- we can add external constraints from the SMF and XLF (named ``the SMF-XLF constraints'' hereafter) into our posterior. This approach was adopted in \citet{Yang18a}. As a start, we revisit the numerical computations of $\phi_{L, \mathrm{mdl}}$ in Equation~\ref{eq: smf_to_xlf}. Again, numerical integrations should be avoided whenever possible, and we hence derive an analytical formula for $\phi_{L, \mathrm{mdl}}$. $\phi_M$ is expressed as a two-component Schechter function in \citet{Wright18}:
\begin{align}
\phi_M=\frac{dn}{d\log M_\star}=\ln10~e^{-\frac{M_\star}{M_c}}\left[\phi_1\left(\frac{M_\star}{M_c}\right)^{\alpha_1+1}+\phi_2\left(\frac{M}{M_c}\right)^{\alpha_2+1}\right],
\end{align}
where $(M_c, \alpha_1, \alpha_2, \phi_1, \phi_2)$ are redshift-dependent parameters. We further define an auxiliary function $\psi$ such that the model-predicted XLF in Equation~\ref{eq: smf_to_xlf} can be simplified as summations of $\psi$ (see below).
\begin{align}
\psi&(\gamma, M_1, M_2, A, \lambda_c; L_\mathrm{X})\nonumber\\
&=\int_{\log M_1}^{\log M_2}A\left(\frac{L_\mathrm{X}}{M_\star\lambda_c}\right)^{-\gamma}\phi_Md\log M_\star\nonumber\\
&=A\left(\frac{L_\mathrm{X}}{M_\star\lambda_c}\right)^{-\gamma}\left[\phi_1\Gamma_\mathrm{GI}\left(\alpha_1+\gamma+1, \frac{M_1}{M_c}, \frac{M_2}{M_c}\right)\right.\nonumber\\
&\left.+\phi_2\Gamma_\mathrm{GI}\left(\alpha_2+\gamma+1, \frac{M_1}{M_c}, \frac{M_2}{M_c}\right)\right],\label{eq: psi}
\end{align}
where $\Gamma_\mathrm{GI}(\zeta, x_1, x_2)=\int_{x_1}^{x_2}t^{\zeta-1}e^{-t}dt$ is the generalized incomplete Gamma function. The contribution of each grid element to the integration in Equation~\ref{eq: smf_to_xlf} is
\begin{align}
\psi_\mathrm{DP}&(A, \lambda_c, \gamma_1, \gamma_2, M_1, M_2; L_\mathrm{X})\nonumber\\
&=\int_{\log M_1}^{\log M_2}p(L_\mathrm{X}/M_\star\mid M_\star, z)\phi_Md\log M_\star\nonumber\\
&=
\begin{cases}
\psi(\gamma_2, M_1, M_2, A, \lambda_c; L_\mathrm{X}), \lambda_c\leq L_\mathrm{X}/M_2\\
\psi(\gamma_2, M_1, L_\mathrm{X}/\lambda_c, A, \lambda_c; L_\mathrm{X})\\
+\psi(\gamma_1, L_\mathrm{X}/\lambda_c, M_2, A, \lambda_c; L_\mathrm{X}), L_\mathrm{X}/M_2<\lambda_c<L_\mathrm{X}/M_1,\\
\psi(\gamma_1, M_1, M_2, A, \lambda_c; L_\mathrm{X}), \lambda_c\geq L_\mathrm{X}/M_1
\end{cases}\label{eq: psi_DP}
\end{align}
Equation~\ref{eq: smf_to_xlf} is thus
\begin{align}
\phi_{L, \mathrm{mdl}}=\sum_{i=1}^{N_M}\psi_\mathrm{DP}&(A_{ij_z}, \lambda_{c,ij_z}, \gamma_{1,ij_z}, \gamma_{2,ij_z}, M_{LB, i}, M_{LB, i+1}; L_\mathrm{X}),\label{eq: phiL_mdl_psiDP}
\end{align}
where $\log M_{LB, i}=\log M_{\star, \mathrm{min}}+(i-1)/N_M\times\log(M_{\star, \mathrm{max}}/M_{\star, \mathrm{min}})$ is the lower bound of the $i^\mathrm{th}$ $M_\star$-grid element, and $j_z$ is the index of the $z$-grid element containing $z$.\par
We then follow the procedure in \citet{Yang18a} to compare $\phi_{L, \mathrm{mdl}}$ and $\phi_{L, \mathrm{obs}}$ in \citet{Ueda14}. $\phi_{L, \mathrm{obs}}$ is evaluated at several $(L_\mathrm{X}, z)$ values, and the number of sources ($n^\mathrm{XLF}$) in \citet{Ueda14} contributing to $\phi_{L, \mathrm{obs}}$ is recorded. Following \citet{Yang18a}, we use the soft-band XLF at $L_\mathrm{X}>10^{43}~\mathrm{erg~s^{-1}}$ in \citet{Ueda14}. Their soft-band XLF has been corrected for obscuration and spans a wider $L_\mathrm{X}$ range compared to their HB XLF, and their soft-band and HB XLFs are also consistent (see Figure~\ref{fig_xlf}). The $L_\mathrm{X}$ cut at $10^{43}~\mathrm{erg~s^{-1}}$ is adopted to avoid contamination from \mbox{X-ray} binaries. The log-likelihood when comparing $\phi_{L, \mathrm{mdl}}$ and $\phi_{L, \mathrm{obs}}$ is
\begin{align}
\ln\mathcal{L}_\mathrm{SMF-XLF}&=\sum_k\ln\mathrm{Pr}\left(\mathrm{Poisson}\left(\frac{\phi_{L, \mathrm{mdl}, k}}{\phi_{L, \mathrm{obs}, k}}n_k^\mathrm{XLF}\right)=n_k^\mathrm{XLF}\right)\nonumber\\
&=\sum_kn_k^\mathrm{XLF}\left[\ln\left(\frac{\phi_{L, \mathrm{mdl}, k}}{\phi_{L, \mathrm{obs}, k}}\right)-\frac{\phi_{L, \mathrm{mdl}, k}}{\phi_{L, \mathrm{obs}, k}}\right]+\mathrm{const}.,\label{eq: lnlike_xlf}\\
\phi_{L, \mathrm{mdl}, k}&=\phi_{L, \mathrm{mdl}}(L_{\mathrm{X}, k}, z_k),
\end{align}
where $k$ runs over all the $L_\mathrm{X}$ and $z$ bins of the observed XLF in \citet{Ueda14}. This term is called the SMF-XLF likelihood in \citet{Yang18a}.\par
\begin{figure*}
\centering
\resizebox{\hsize}{!}{\includegraphics{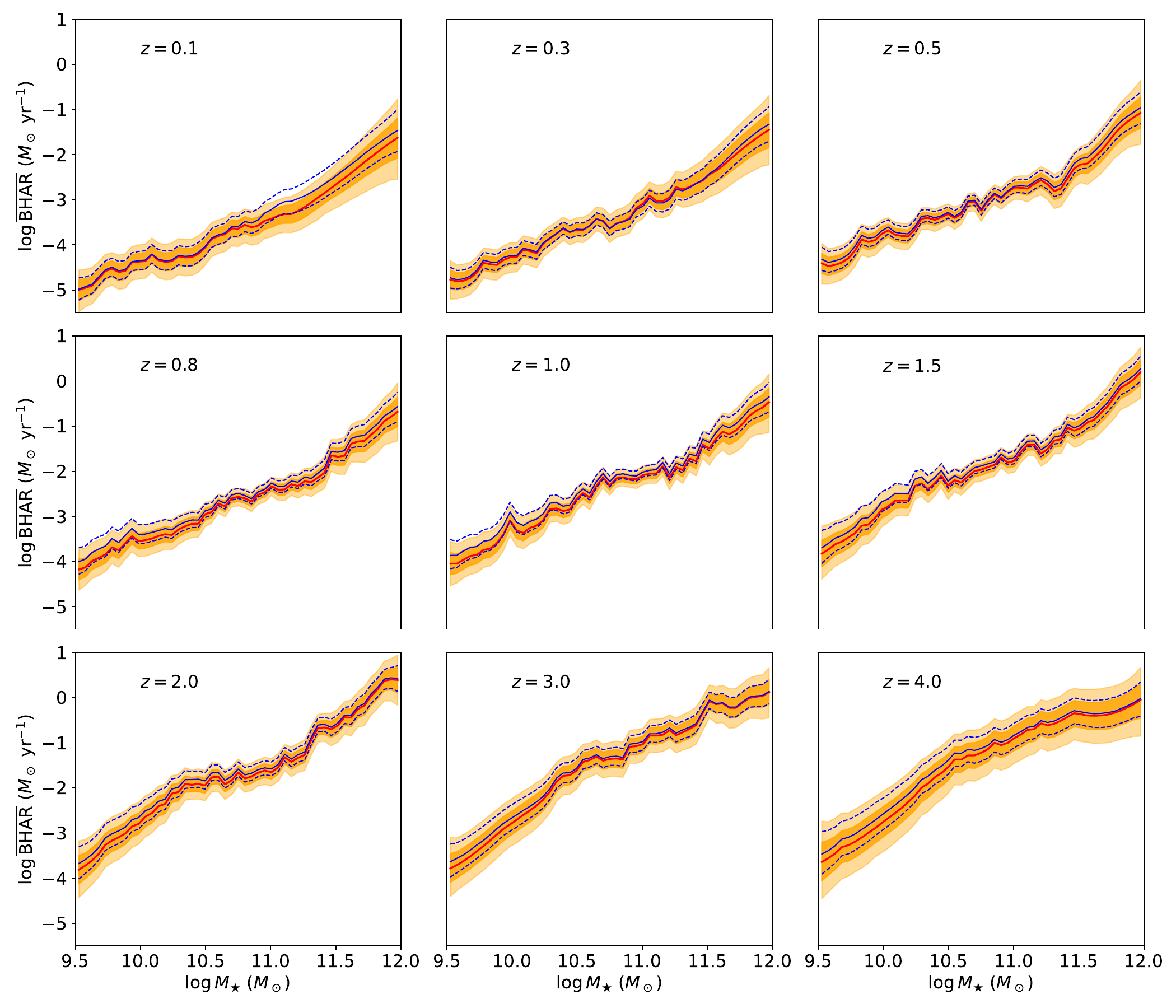}}
\caption{$\overline{\mathrm{BHAR}}$ as a function of $M_\star$ at several redshifts. The red curves are our median $\overline{\mathrm{BHAR}}$ with the SMF-XLF constraints added, and the orange shaded regions represent the corresponding $1\sigma$ and $2\sigma$ uncertainty ranges. The blue curves are our median $\overline{\mathrm{BHAR}}$ and $1\sigma$ uncertainties without the SMF-XLF constraints.}
\label{fig_bharcompare_withxlf}
\end{figure*}

To add the SMF-XLF constraints, Equation~\ref{eq: lnpost} should be modified as follows.
\begin{align}
\ln\mathcal{P}&=\sum_\mathrm{field}\ln\mathcal{L}+\ln\mathcal{L}_\mathrm{SMF-XLF}+\ln\pi_\mathrm{cont}.\label{eq: lnpost_smf_xlf}
\end{align}
Its gradient is presented in Appendix~\ref{append: gradient_smf_xlf} for HMC sampling. We then sample the above posterior with HMC and present the resulting $\overline{\mathrm{BHAR}}$ in Figure~\ref{fig_bharcompare_withxlf}. The $\overline{\mathrm{BHAR}}$ curves with or without the SMF-XLF constraints are largely consistent with a small ($<1\sigma$) difference. This is expected because Figure~\ref{fig_xlf} shows that our $\overline{\mathrm{BHAR}}$ without the SMF-XLF constraints leads to consistent XLFs with those in \citet{Ueda14}.\par

Although there is a good consistency after adding the SMF-XLF constraints in our case, extra cautions are generally needed. The SMF and XLF taken from other literature works usually involve inherent assumptions about their parametric forms. When putting the SMF and XLF into our posterior, we will inevitably ``absorb'' these assumptions. Besides, the original data used to measure the XLF may overlap with one's dataset, especially given that the \mbox{X-ray} data in GOODS-S are also necessary to constrain the XLF at low-$L_\mathrm{X}$ and/or high-$z$. Such an overlap causes double counting of the involved sources. Especially, more considerations would be needed if the posterior is dominated by the SMF-XLF constraints.

\subsection{Comparison with Previous Works}
\label{sec: compare_literature}
Figure~\ref{fig_plambda_literature} compares our $p(\lambda\mid M_\star, z)$ with some representative results in previous literature. \citet[black solid lines in Figure~\ref{fig_plambda_literature}]{Aird12} used a single power-law to fit $p(L_\mathrm{X}\mid M_\star, z)$ at $z<1$, which is converted to a single power-law $p(\lambda\mid M_\star, z)$ in Figure~\ref{fig_plambda_literature}. The single power-law curves broadly follow our double power-law ones, and the single power-law index lies within the range between $\gamma_1$ and $\gamma_2$. This indicates that a single power-law model can serve well as the first-order approximation of $p(\lambda\mid M_\star, z)$, as has been widely adopted in other works (e.g., \citealt{Bongiorno12, Wang17, Birchall20, Birchall23, Zou23}), especially when the data are limited. However, the real $p(\lambda\mid M_\star, z)$ is more complicated, and a double power-law model can return better characterizations. As Figure~\ref{fig_plambda_literature} shows, the binned $p(\lambda\mid M_\star, z)$ estimators generally do not show systematic deviations from our double power-law curves (e.g., no further breaks are visible), and thus a double power-law model is sufficient to capture the main structures of $p(\lambda\mid M_\star, z)$ at $\lambda>\lambda_\mathrm{min}$.\par
\citet{Bongiorno16} and \citet{Yang18a} adopted a double power-law model similar to ours, and we plot their results as the dash-dotted and dashed lines in Figure~\ref{fig_plambda_literature}, respectively. Our $p(\lambda\mid M_\star, z)$ curves are nearly identical to those in \citet{Yang18a} at $10\lesssim\log M_\star\lesssim11.5$ and $1\lesssim z\lesssim2.5$ but begin diverging in other parameter ranges. In the lowest-mass bin ($9.5<\log M_\star<10$), our $p(\lambda\mid M_\star, z)$ is still similar to those in \citet{Yang18a} at $\log\lambda\lesssim33.5$ but is lower than theirs at higher $\lambda$. In the highest-mass bin ($11.5<\log M_\star<12$), our $p(\lambda\mid M_\star, z)$ is larger at $z\lesssim2$ and smaller at $z\gtrsim2$ than for \citet{Yang18a}. It should be noted that these parameter regions with noticeable $p(\lambda\mid M_\star, z)$ differences generally have limited data and are far away from the bulk of other data points, and the results in these regions are subject to large uncertainties. For \citet{Bongiorno16}, their $p(\lambda\mid M_\star, z)$ is similar to ours at $10\lesssim\log M_\star\lesssim11.5$ and $1.5\lesssim z\lesssim2$ but has a much steeper low-$\lambda$ power-law index at $z<1.5$. Two reasons may be responsible for the difference -- the data used in \citet{Bongiorno16} are not sufficiently deep to effectively probe the low-$\lambda$ regime; their model always fixes the breakpoint at $\log\lambda=33.8$ when $M_\star=10^{11}~M_\odot$, while our results suggest that the breakpoint tends to become smaller as redshift decreases.\par
\citet{Georgakakis17} and \citet{Aird18} adopted nonparametric methods to measure $p(\lambda\mid M_\star, z)$ without assuming a double power-law form. Our results show good agreement with theirs, especially in regimes well-covered by the data, suggesting that a double power-law is indeed a good approximation of $p(\lambda\mid M_\star, z)$. Nonetheless, some differences are worth noting. At $\log\lambda\gtrsim34$ where the data become limited, the $p(\lambda\mid M_\star, z)$ in \citet{Aird18} tends to be flatter than ours, while that in \citet{Georgakakis17} tends to be steeper than ours. This high-$\lambda$ regime is highly uncertain and subject to the adopted methodology -- for instance, the prior adopted in \citet{Aird18} prefers a flatter slope at high $\lambda$. Another feature is that the $p(\lambda\mid M_\star, z)$ in \citet{Aird18} sometimes shows downward bending at $\log\lambda\approx32-33$, while that in \citet{Georgakakis17} does not show a clear bending, although the large uncertainty may be responsible for the lack of bending. In principle, a downward bending at some low $\lambda$ is expected; otherwise, $p(\lambda\mid M_\star, z)$ would diverge. Such bending can also be seen in \citet{Georgakakis17}, but below $\log\lambda_\mathrm{min}=31.5$ (see, e.g., their Figure~7). Our double power-law model is unable to capture this feature, and Figure~\ref{fig_plambda_literature} shows that the bending in \citet{Aird18} mainly becomes prominent at high redshift ($z\gtrsim3$).\par
\begin{figure}
\centering
\resizebox{\hsize}{!}{\includegraphics{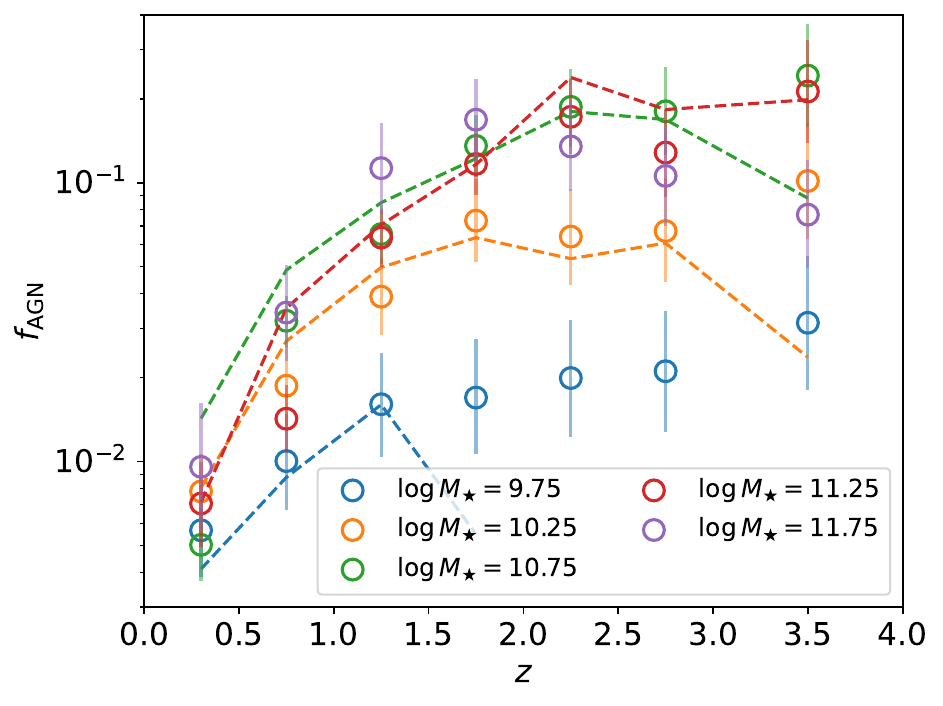}}
\caption{$f_\mathrm{AGN}$ evaluated at several $(M_\star, z)$ values versus $z$. Our results are plotted as open circles with $1\sigma$ error bars, where different colors represent different $M_\star$. The dashed lines are those in \citet{Aird18}.}
\label{fig_fduty}
\end{figure}

Another metric that can be measured from $p(\lambda\mid M_\star, z)$ is the fraction of galaxies hosting accreting SMBHs above a given $\lambda$ threshold ($\lambda_\mathrm{thres}$), as calculated below.
\begin{align}
f_\mathrm{AGN}(\lambda>\lambda_\mathrm{thres})=\int_{\log\lambda_\mathrm{thres}}^{+\infty}p(\lambda\mid M_\star, z)d\log\lambda.
\end{align}
For a consistent comparison with \citet{Aird18}, we adopt the same $\lambda_\mathrm{thres}=32$ as theirs. We calculate $f_\mathrm{AGN}$ at several $(M_\star, z)$ values and plot the results in Figure~\ref{fig_fduty}. Our results generally agree well with those in \citet{Aird18} and follow similar evolutionary trends with respect to $M_\star$ and $z$. At $\log M_\star\gtrsim10$, $f_\mathrm{AGN}$ increases with $z$ up to $z\approx1.5-2$ and reaches a plateau at higher redshift; while for less-massive galaxies, the redshift evolution is weaker. At low redshift ($z\lesssim0.5$), $f_\mathrm{AGN}$ is similar regardless of $M_\star$, and this conclusion can be further extended down to $\log M_\star<9.5$, as \citet{Zou23} showed that the $\lambda$-based $f_\mathrm{AGN}$ in the dwarf-galaxy population in this redshift range is also similar to $f_\mathrm{AGN}$ in massive galaxies. At higher redshift ($z\gtrsim1$), the dependence of $f_\mathrm{AGN}$ on $M_\star$ becomes more apparent due to $M_\star$-dependent redshift evolution rates of $f_\mathrm{AGN}$, and there is a positive correlation between $f_\mathrm{AGN}$ and $M_\star$ at $\log M_\star\lesssim10.5$. However, for massive galaxies with $\log M_\star\gtrsim10.5$, $f_\mathrm{AGN}$ nearly does not depend on $M_\star$. A full physical explanation of these complicated correlations between $f_\mathrm{AGN}$ and $(M_\star, z)$ will require further detailed analyses of $p(\lambda\mid M_\star, z)$ with at least partially physically driven modeling, and we leave these analyses for future work.\par

\begin{figure*}
\centering
\resizebox{\hsize}{!}{\includegraphics{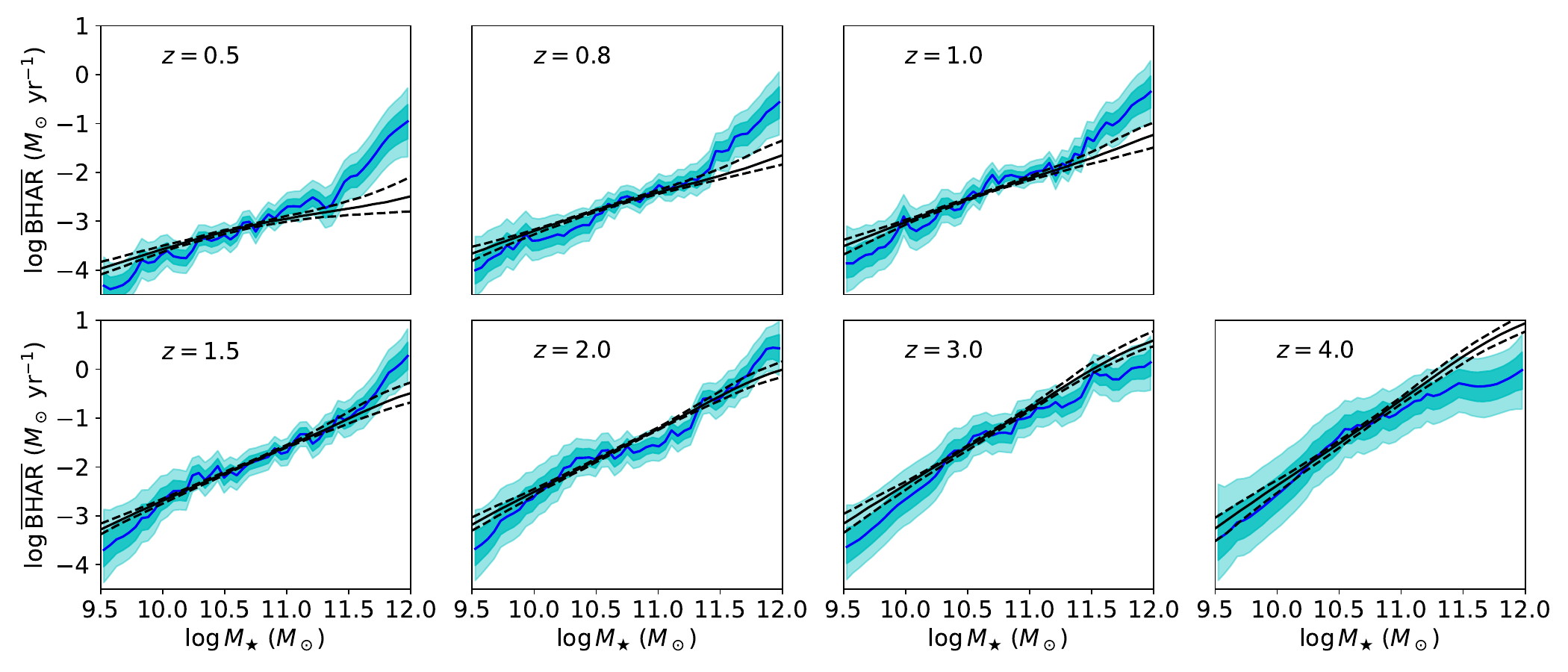}}
\caption{The comparison of our $\overline{\mathrm{BHAR}}$ with those in \citet{Yang18a}. The blue curves are our median $\overline{\mathrm{BHAR}}$, and the cyan shaded regions represent the corresponding $1\sigma$ and $2\sigma$ uncertainty ranges. The $\overline{\mathrm{BHAR}}$ and the corresponding $1\sigma$ uncertainty in \citet{Yang18a} are plotted as the black curves.}
\label{fig_bharcompare}
\end{figure*}

We further compare our $\overline{\mathrm{BHAR}}$ with those from \citet{Yang18a} in Figure~\ref{fig_bharcompare}. Our median relation is largely similar to theirs, but some subtle differences exist. Our low-mass $\overline{\mathrm{BHAR}}$ at $\log M_\star\lesssim10$ is slightly smaller across all redshifts, though not very significant. Our high-mass $\overline{\mathrm{BHAR}}$ at $\log M_\star\gtrsim11.5$ differs the most with \citet{Yang18a}, and ours tends to be smaller at $z\gtrsim3$ while being larger at $z\lesssim2$. These differences originate from different $p(\lambda\mid M_\star, z)$, as discussed earlier in this section. As shown in Figure~\ref{fig_plambda_literature}, our low-mass $p(\lambda\mid M_\star, z)$ is smaller than for \citet{Yang18a} only at high $\lambda$, and thus the low-mass $\overline{\mathrm{BHAR}}$ difference is small. Our high-mass $p(\lambda\mid M_\star, z)$ at $\log M_\star\gtrsim11.5$, instead, shows a redshift-dependent difference in the normalization. Nevertheless, the uncertainties in these extreme regimes are large, and they are also subject to model choices. Dedicated analyses of these extreme-mass sources with deeper or wider data may be necessary to further pin down the uncertainty. Another important difference is that the $\overline{\mathrm{BHAR}}-M_\star$ relation in \citet{Yang18a} flattens at low redshift, but ours do not show such a trend. Therefore, the $\overline{\mathrm{BHAR}}$ in \citet{Yang18a} is less reliable at $z\lesssim0.8$, as they noted; if their relation is further extrapolated below $z=0.5$, their $\overline{\mathrm{BHAR}}-M_\star$ relation would become flat and is thus unphysical. Our $\overline{\mathrm{BHAR}}$ uncertainties are also considerably larger than those in \citet{Yang18a}, even though we used more data. This is because \citet{Yang18a} adopted a parametric modeling method, which includes strong a priori assumptions. In contrast, this work minimizes such assumptions, and thus the fitted uncertainties reflect those directly inherited from the data.\par

\subsection{Star-forming versus Quiescent Galaxies}
\label{sec: sfvq}
Star-forming galaxies generally have stronger AGN activity than quiescent galaxies (e.g., \citealt{Aird18, Aird19}). We hence examine if star-forming and quiescent galaxies have the same $\overline{\mathrm{BHAR}}$ in this section.\par
\begin{figure*}
\centering
\resizebox{\hsize}{!}{\includegraphics{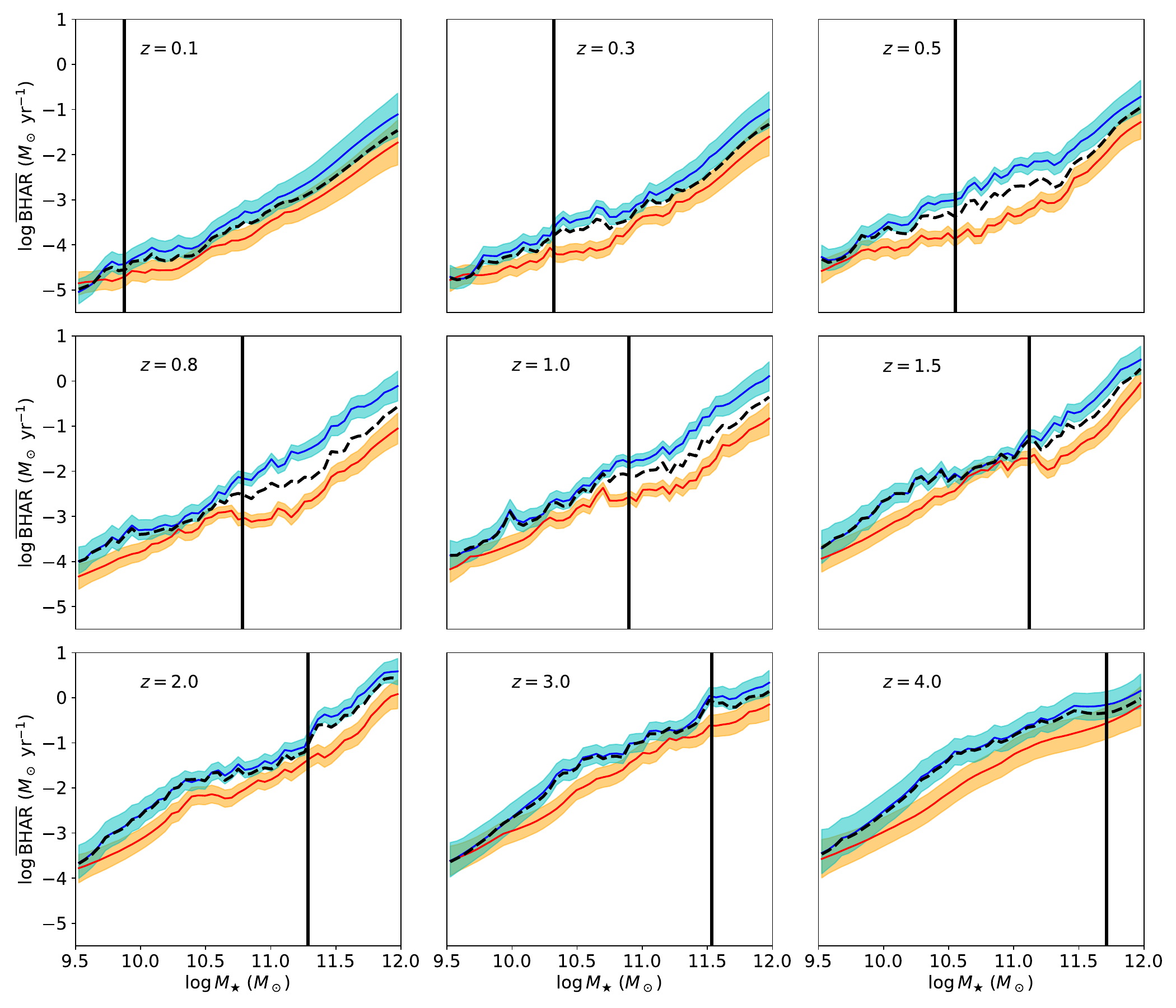}}
\caption{$\overline{\mathrm{BHAR}}$ for star-forming (blue) and quiescent (red) galaxies. The shaded regions represent $1\sigma$ uncertainty ranges. The black dashed curves are the $\overline{\mathrm{BHAR}}$ with all the galaxies included, i.e., those in Figure~\ref{fig_bharresult}. The vertical black lines mark the maximum $M_\star$ values where star-forming and quiescent galaxies can be reliably classified at the corresponding $z$ \citep{Cristello24}. Star-forming galaxies have larger $\overline{\mathrm{BHAR}}$.}
\label{fig_bharcompare_sfq}
\end{figure*}

To separate star-forming and quiescent galaxies, we adopt the star-forming main sequence (MS) in \citet{Popesso23} and define quiescent galaxies as those lying at least 1~dex below the MS; the remaining galaxies are star-forming ones. Since the star-forming and quiescent subpopulations do not individually follow the SMF and XLF, we do not apply the SMF-XLF constraints as in Section~\ref{sec: addxlf}. We measure their $\overline{\mathrm{BHAR}}$ and present the results in Figure~\ref{fig_bharcompare_sfq}. The $\overline{\mathrm{BHAR}}$ of both star-forming and quiescent galaxies increases with $M_\star$ and $z$. When comparing the $\overline{\mathrm{BHAR}}$ of these two subpopulations, star-forming galaxies generally have larger $\overline{\mathrm{BHAR}}$, suggesting that star-forming galaxies indeed host more active SMBHs, possibly due to more available cold gas for both star formation and SMBH accretion. The $\overline{\mathrm{BHAR}}$ difference between the two populations also depends on $M_\star$ and $z$. At $\log M_\star\lesssim10.5$, the difference is generally small across most of the redshift range. At higher mass, the difference is small at low redshift but becomes apparent when $z$ increases to 1 and further decreases at higher redshift. There is also tentative evidence suggesting that the redshift at which the difference reaches its peak might also shift with $M_\star$, with the peak of the $\overline{\mathrm{BHAR}}$ difference of higher-mass galaxies occurring at higher redshift.\par

One caveat that should be noted is that our results depend on the classification between star-forming and quiescent galaxies. Such a classification is more reliable at $M_\star\lesssim10^{10.5}~M_\odot$, but it may become sensitive to the adopted method at higher $M_\star$ and lower $z$ (e.g., \citealt{Donnari19}). \citet{Cristello24} show that the star-forming and quiescent subpopulations cannot be safely defined for massive galaxies, and \citet{Feldmann17} also argued that the bimodal separation is not necessarily appropriate. The proposed redshift-dependent maximum $M_\star$ values for reliable classifications in \citet{Cristello24} can be well described by the following equation:
\begin{align}
\log M_\star=10.65+0.81\log(z)+0.83\log(1+z),\label{eq: safemass}
\end{align}
and they are explicitly plotted in Figure~\ref{fig_bharcompare_sfq}. We also plot the $\overline{\mathrm{BHAR}}$ of the whole population in Figure~\ref{fig_bharcompare_sfq}, and it is similar to the star-forming $\overline{\mathrm{BHAR}}$ below the $M_\star$ threshold in Equation~\ref{eq: safemass} and becomes more in the middle between the star-forming and quiescent $\overline{\mathrm{BHAR}}$ with rising $M_\star$. Therefore, Equation~\ref{eq: safemass} can also serve as an approximate threshold of whether the contribution of the SMBH growth in quiescent galaxies to the overall SMBH growth is important.\par
Our results suggest that the $\overline{\mathrm{BHAR}}(M_\star, z)$ function may also depend on SFR, with star-forming galaxies hosting enhanced AGN activity (e.g., \citealt{Aird18, Aird19, Birchall23}). However, such a dependence is only secondary \citep{Yang17}, and SFR is usually more challenging to measure and more subject to confusion with AGN emission. Still, more physical insights can be gained by incorporating SFR-based parameters, especially when probing $p(\lambda\mid M_\star, z)$ instead of $\overline{\mathrm{BHAR}}$ \citep{Aird18}. We leave further analyses on including SFR into the $\overline{\mathrm{BHAR}}(M_\star, z)$ function to the future, in which different classification schemes from binary (star-forming versus quiescent) up to four categories (starburst, star-forming, transitioning, and quiescent) will be explored.

\section{Summary and Future Work}
\label{sec: summary}
In this work, we mapped $\overline{\mathrm{BHAR}}$ as a function of $(M_\star, z)$ over the vast majority of cosmic time, and our main results are summarized as follows:
\begin{enumerate}
\item{
We compiled an unprecedentedly large sample from nine fields -- CANDELS (including GOODS-S, GOODS-N, EGS, and UDS), the LSST DDFs (including COSMOS, ELAIS-S1, W-CDF-S, and XMM-LSS), and eFEDS. These fields include both deep, small-area surveys and shallow, large-area ones. The former provide rich information in the high-$z$, low-$M_\star$, and/or low-$\lambda$ regime, while the latter provide complementary information in the low-$z$, high-$M_\star$, and/or high-$\lambda$ regime. Therefore, our sample can effectively constrain $\overline{\mathrm{BHAR}}$ across a large range of the relevant parameter space. See Section~\ref{sec: data}.
}
\item{
We developed a semiparametric Bayesian method to measure $\overline{\mathrm{BHAR}}$, where a double power-law model with respect to $\lambda$ is used to measure $p(\lambda\mid M_\star, z)$, and the relevant parameters nonparametrically depend on $(M_\star, z)$. This method has two main advantages. It avoids the ``extrapolation'' of parameters from well-populated regions in the parameter space to less-populated regions. It also adopts much weaker assumptions than parametric methods, enabling more flexible constraints and more representative fitting uncertainties from the data. See Section~\ref{sec: plambda}.
}
\item{
We sampled $p(\lambda\mid M_\star, z)$ and measured $\overline{\mathrm{BHAR}}$ at $10^{9.5}<M_\star<10^{12}~M_\odot$ and $z<4$. We have verified the fitting quality by comparing our model $p(\lambda\mid M_\star, z)$ and the corresponding binned estimators and also by comparing our inferred XLF with the observed one. We showed that $\overline{\mathrm{BHAR}}$ increases with both $M_\star$ and $z$. Our $\overline{\mathrm{BHAR}}$ measurements are largely consistent with those in \citet{Yang18a} at $z\gtrsim0.8$, and we also, for the first time, provide reasonable constraints upon $\overline{\mathrm{BHAR}}$ at lower redshift ($z\lesssim0.5$). See Sections~\ref{sec: hmcsample} and \ref{sec: bhar}.
}
\item{
We measured $\overline{\mathrm{BHAR}}$ for both star-forming and, for the first time, quiescent galaxies. Both groups show $\overline{\mathrm{BHAR}}$ increases with $M_\star$ and $z$, and the star-forming $\overline{\mathrm{BHAR}}$ is generally larger than or at least comparable to the quiescent $\overline{\mathrm{BHAR}}$ across the whole $(M_\star, z)$ plane. See Section~\ref{sec: sfvq}.
}
\end{enumerate}
It should be noted that, besides $\overline{\mathrm{BHAR}}$, our $p(\lambda\mid M_\star, z)$ parameter maps in Figure~\ref{fig_thetamap} also contain rich information, and we release $p(\lambda\mid M_\star, z)$ and the corresponding parameter maps and $\overline{\mathrm{BHAR}}$ maps in \url{https://doi.org/10.5281/zenodo.10729248}. As first examples, we have briefly and phenomenologically discussed different scientific questions in Sections~\ref{sec: compare_literature} and \ref{sec: sfvq}, which justified that our results can reveal interesting dependences of SMBH growth on the galaxy population.\par
Figure~\ref{fig_plambda_literature} visually illustrates that $p(\lambda\mid M_\star, z)$ evolves over $(M_\star, z)$. Observationally, it is still unclear what the exact evolution pattern is, let alone the physics driving such an evolution. It is also unknown from a theoretical perspective because no simulations appear to produce consistent evolution patterns of $p(\lambda\mid M_\star, z)$ with the observed ones (e.g., \citealt{Habouzit22}). It even complicates matters further that $p(\lambda\mid M_\star, z)$ may evolve differently for star-forming and quiescent galaxies, as proposed in a phenomenological scenario in \citet{Aird18}. We leave detailed analyses of the $p(\lambda\mid M_\star, z)$ evolution to a subsequent future work. We will first identify the qualitative evolution pattern of the dependence of $p(\lambda\mid M_\star, z)$ on $M_\star$ and $z$ for different galaxy populations and then develop a quantitative, parametric model to depict the identified evolution pattern. With the clearly understood $p(\lambda\mid M_\star, z)$, we will address the following scientific questions. Is the broad decline in SMBH growth below $z\approx1$ due to the shift of accretion activity to smaller galaxies or reduction of the typical $\lambda$? How large is the AGN ``duty cycle'', which is an integration of $p(\lambda\mid M_\star, z)$, in different galaxy populations? Does $M_\star$ modulate the duty cycle or modulate the typical outburst luminosity in the AGN phase? Is there any difference in the SMBH feeding in star-forming and quiescent galaxies?

\acknowledgments
We thank the anonymous referee for constructive suggestions and comments. We thank Nathan Cristello, Joel Leja, and Zhenyuan Wang for helpful discussions. FZ, ZY, and WNB acknowledge financial support from NSF grant AST-2106990, Chandra X-ray Center grant AR1-22006X, the Penn State Eberly Endowment, and Penn State ACIS Instrument Team Contract SV4-74018 (issued by the Chandra X-ray Center, which is operated by the Smithsonian Astrophysical Observatory for and on behalf of NASA under contract NAS8-03060). GY acknowledges funding from the Netherlands Research School for Astronomy (NOVA). The Chandra ACIS Team Guaranteed Time Observations (GTO) utilized were selected by the ACIS Instrument Principal Investigator, Gordon P. Garmire, currently of the Huntingdon Institute for X-ray Astronomy, LLC, which is under contract to the Smithsonian Astrophysical Observatory via Contract SV2-82024.

\appendix
\section{Gradient of the Posterior}
\label{append: gradient}
This Appendix presents the gradient of our posterior in Equation~\ref{eq: lnpost}. We found that, at least in our case, analytical differentiation enables a much higher computational speed and/or less memory compared with other differentiation algorithms. We thus adopt the analytically-derived gradient and directly present the derivation results below.\par
First, the partial derivatives of $I(\gamma, \lambda_1, \lambda_2, A, \lambda_c; M_\star, z)$ in Equation~\ref{eq: I} are
\begin{align}
\frac{\partial I}{\partial A}=\frac{I}{A}
\end{align}
\begin{align}
\frac{\partial I}{\partial\lambda_c}=\frac{\gamma}{\lambda_c}I
\end{align}
\begin{align}
\frac{\partial I}{\partial\gamma}&=-\frac{A}{2\gamma\ln10}\left[\ln\left(\frac{\lambda_1}{\lambda_c}\right)+\frac{1}{\gamma}\right]\left(\frac{\lambda_1}{\lambda_c}\right)^{-\gamma}\left[\mathrm{erf}(x_1)+1\right]\nonumber\\
&+\frac{A}{2\gamma\ln10}\left[\ln\left(\frac{\lambda_2}{\lambda_c}\right)+\frac{1}{\gamma}\right]\left(\frac{\lambda_2}{\lambda_c}\right)^{-\gamma}\left[\mathrm{erf}(x_2)+1\right]\nonumber\\
&+\frac{A}{2\gamma\ln10}\left(\frac{10^{a-\frac{\gamma\ln10}{4b^2}}}{\lambda_cM_\star\eta}\right)^{-\gamma}\left[\ln\left(\frac{10^a}{\lambda_cM_\star\eta}\right)-\frac{\gamma(\ln10)^2}{2b^2}+\frac{1}{\gamma}\right]\times\nonumber\\
&\left[\mathrm{erf}\left(x_1+\frac{\gamma\ln10}{2b}\right)-\mathrm{erf}\left(x_2+\frac{\gamma\ln10}{2b}\right)\right]\nonumber\\
&-\frac{A}{2\sqrt{\pi}\gamma b}\left(\frac{10^{a-\frac{\gamma\ln10}{4b^2}}}{\lambda_cM_\star\eta}\right)^{-\gamma}\times\nonumber\\
&\left[\exp\left(-\left(x_1+\frac{\gamma\ln10}{2b}\right)^2\right)-\exp\left(-\left(x_2+\frac{\gamma\ln10}{2b}\right)^2\right)\right]
\end{align}
\begin{align}
\frac{\partial I}{\partial\lambda_k}&=(2k-3)\times\nonumber\\
&\Biggl\{\frac{A}{2\lambda_k\ln10}\left(\frac{\lambda_k}{\lambda_c}\right)^{-\gamma}\left[\mathrm{erf}(x_k)+1\right]\nonumber\\
&-\frac{Ab}{\sqrt{\pi}(\ln10)^2\gamma\lambda_k}\left(\frac{\lambda_k}{\lambda_c}\right)^{-\gamma}\exp\left(-x_k^2\right)\nonumber\\
&+\frac{Ab}{\sqrt{\pi}(\ln10)^2\gamma\lambda_k}\left(\frac{10^{a-\frac{\gamma\ln10}{4b^2}}}{\lambda_cM_\star\eta}\right)^{-\gamma}\exp\left(-\left(x_k+\frac{\gamma\ln10}{2b}\right)^2\right)\Biggl\}\nonumber\\
&(k=1, 2)
\end{align}
Defining $\ln p(\lambda; A, \lambda_c, \gamma_1, \gamma_2)$ as $\ln p(\lambda\mid M_\star, z)$, its partial derivatives are
\begin{align}
\frac{\partial\ln p}{\partial A}&=\frac{1}{A}
\end{align}
\begin{align}
\frac{\partial\ln p}{\partial\lambda_c}&=
\begin{cases}
\frac{\gamma_1}{\lambda_c}, \lambda<\lambda_c\\
\frac{\gamma_2}{\lambda_c}, \lambda>\lambda_c
\end{cases}
\end{align}
\begin{align}
\frac{\partial\ln p}{\partial\gamma_1}&=
\begin{cases}
-\ln\left(\frac{\lambda}{\lambda_c}\right), \lambda<\lambda_c\\
0, \lambda>\lambda_c
\end{cases}
\end{align}
\begin{align}
\frac{\partial\ln p}{\partial\gamma_2}&=
\begin{cases}
0, \lambda<\lambda_c,\\
-\ln\left(\frac{\lambda}{\lambda_c}\right), \lambda>\lambda_c
\end{cases}
\end{align}
$\nabla\ln\mathcal{L}$ corresponding to Equation~\ref{eq: lnlike_final} can then be expressed as follows.
\begin{align}
\frac{\partial\ln\mathcal{L}}{\partial A_{ij}}&=-n_{ij}^\mathrm{gal}\left[\frac{\partial I}{\partial A}(\gamma_{1, ij}, \lambda_\mathrm{min}, \lambda_{c, ij}, A_{ij}, \lambda_{c, ij}; M_{\star, i}, z_j)\right.\nonumber\\
&\left.+\frac{\partial I}{\partial A}(\gamma_{2, ij}, \lambda_{c, ij}, +\infty, A_{ij}, \lambda_{c, ij}; M_{\star, i}, z_j)\right]\nonumber\\
&+\sum_{s=1}^{n_{ij}^\mathrm{AGN}}\frac{\partial\ln p}{\partial A}(\lambda_s; A_{ij}, \lambda_{c, ij}, \gamma_{1, ij}, \gamma_{2, ij})
\end{align}
\begin{align}
\frac{\partial\ln\mathcal{L}}{\partial \lambda_{c, ij}}&=-n_{ij}^\mathrm{gal}\left[\frac{\partial I}{\partial\lambda_2}(\gamma_{1, ij}, \lambda_\mathrm{min}, \lambda_{c, ij}, A_{ij}, \lambda_{c, ij}; M_{\star, i}, z_j)\right.\nonumber\\
&+\frac{\partial I}{\partial\lambda_c}(\gamma_{1, ij}, \lambda_\mathrm{min}, \lambda_{c, ij}, A_{ij}, \lambda_{c, ij}; M_{\star, i}, z_j)\nonumber\\
&+\frac{\partial I}{\partial\lambda_1}(\gamma_{2, ij}, \lambda_{c, ij}, +\infty, A_{ij}, \lambda_{c, ij}; M_{\star, i}, z_j)\nonumber\\
&\left.+\frac{\partial I}{\partial\lambda_c}(\gamma_{2, ij}, \lambda_{c, ij}, +\infty, A_{ij}, \lambda_{c, ij}; M_{\star, i}, z_j)\right]\nonumber\\
&+\sum_{s=1}^{n_{ij}^\mathrm{AGN}}\frac{\partial\ln p}{\partial\lambda_c}(\lambda_s; A_{ij}, \lambda_{c, ij}, \gamma_{1, ij}, \gamma_{2, ij})
\end{align}
\begin{align}
\frac{\partial\ln\mathcal{L}}{\partial\gamma_{k, ij}}&=-n_{ij}^\mathrm{gal}\frac{\partial I}{\partial\gamma}(\gamma_{k, ij}, \lambda_\mathrm{min}, \lambda_{c, ij}, A_{ij}, \lambda_{c, ij}; M_{\star, i}, z_j)\nonumber\\
&+\sum_{s=1}^{n_{ij}^\mathrm{AGN}}\frac{\partial\ln p}{\partial\gamma_k}(\lambda_s; A_{ij}, \lambda_{c, ij}, \gamma_{1, ij}, \gamma_{2, ij})\nonumber\\
&(k=1, 2)
\end{align}\par
The partial derivatives of $\ln\pi_\mathrm{cont}$ in Equation~\ref{eq: lnprior} are
\begin{align}
\frac{\partial\ln\pi_\mathrm{cont}}{\partial X_{ij}}=\frac{N_M(X_{i-1, j}+X_{i+1, j}-2X_{ij})}{\sigma_X^2}+\frac{N_z(X_{i, j-1}+X_{i, j+1}-2X_{ij})}{\sigma_X^2},
\end{align}
in which $X$ denotes each one of $(\log A, \log\lambda_c, \gamma_1, \gamma_2)$, and we define $X_{0j}\equiv X_{1j}$, $X_{N_M+1, j}\equiv X_{N_M, j}$, $X_{i0}\equiv X_{i1}$, and $X_{i, N_z+1}\equiv X_{i, N_z}$ to incorporate $X$'s at the boundary.\par
The gradient of the log-posterior in Equation~\ref{eq: lnpost} is thus
\begin{align}
\nabla\ln\mathcal{P}=\sum_\mathrm{field}\nabla\ln\mathcal{L}+\nabla\ln\pi_\mathrm{cont}.\label{eq: nablalnpost}
\end{align}
When transforming the parameter space, the gradient of the corresponding Jacobian should also be added.

\section{Gradient of the Posterior with the SMF-XLF Constraints Added}
\label{append: gradient_smf_xlf}
This Appendix presents the gradient of our posterior after adding the SMF-XLF constraints in Equation~\ref{eq: lnpost_smf_xlf}. First, the partial derivatives of $\psi(\gamma, M_1, M_2, A, \lambda_c; L_\mathrm{X})$ in Equation~\ref{eq: psi} are
\begin{align}
\frac{\partial\psi}{\partial A}=\frac{\psi}{A}
\end{align}
\begin{align}
\frac{\partial\psi}{\partial\lambda_c}=\frac{\gamma\psi}{\lambda_c}
\end{align}
\begin{align}
\frac{\partial\psi}{\partial\gamma}&=A\left(\frac{L_\mathrm{X}}{M_c\lambda_c}\right)^{-\gamma}\phi_1\Biggl[\frac{\partial\Gamma_\mathrm{GI}}{\partial\zeta}\left(\alpha_1+\gamma+1, \frac{M_1}{M_c}, \frac{M_2}{M_c}\right)\nonumber\\
&-\ln\left(\frac{L_\mathrm{X}}{M_c\lambda_c}\right)\Gamma_\mathrm{GI}\left(\alpha_1+\gamma+1, \frac{M_1}{M_c}, \frac{M_2}{M_c}\right)\Biggl]\nonumber\\
&+A\left(\frac{L_\mathrm{X}}{M_c\lambda_c}\right)^{-\gamma}\phi_2\Biggl[\frac{\partial\Gamma_\mathrm{GI}}{\partial\zeta}\left(\alpha_2+\gamma+1, \frac{M_1}{M_c}, \frac{M_2}{M_c}\right)\nonumber\\
&-\ln\left(\frac{L_\mathrm{X}}{M_c\lambda_c}\right)\Gamma_\mathrm{GI}\left(\alpha_2+\gamma+1, \frac{M_1}{M_c}, \frac{M_2}{M_c}\right)\Biggl]
\end{align}
\begin{align}
\frac{\partial\psi}{\partial M_k}&=(2k-3)\frac{A}{M_c}\left(\frac{L_\mathrm{X}}{M_c\lambda_c}\right)^{-\gamma}e^{-\frac{M_k}{M_c}}\times\nonumber\\
&\left[\phi_1\left(\frac{M_k}{M_c}\right)^{\alpha_1+\gamma}+\phi_2\left(\frac{M_k}{M_c}\right)^{\alpha_2+\gamma}\right]~(k=1, 2)
\end{align}
where $\frac{\partial\Gamma_\mathrm{GI}}{\partial\zeta}(\zeta, x_1, x_2)$ is the partial derivative relative to the first argument of $\Gamma_\mathrm{GI}(\zeta, x_1, x_2)$. The partial derivatives of $\psi_\mathrm{DP}(A, \lambda_c, \gamma_1, \gamma_2, M_1, M_2; L_\mathrm{X})$ in Equation~\ref{eq: psi_DP} are
\begin{align}
\frac{\partial\psi_\mathrm{DP}}{\partial A}=\frac{\psi_\mathrm{DP}}{A}
\end{align}
\begin{align}
\frac{\partial\psi_\mathrm{DP}}{\partial\lambda_c}=
\begin{cases}
\frac{\partial\psi}{\partial\lambda_c}(\gamma_2, M_1, M_2, A, \lambda_c), \lambda_c<\frac{L_\mathrm{X}}{M_2}\\
\frac{\partial\psi}{\partial\lambda_c}(\gamma_2, M_1, \frac{L_\mathrm{X}}{\lambda_c}, A, \lambda_c)+\frac{\partial\psi}{\partial\lambda_c}(\gamma_1, \frac{L_\mathrm{X}}{\lambda_c}, M_2, A, \lambda_c)\\
-\frac{L_\mathrm{X}}{\lambda_c^2}\left[\frac{\partial\psi}{\partial M_2}(\gamma_2, M_1, \frac{L_\mathrm{X}}{\lambda_c}, A, \lambda_c)\right.\\
\left.+\frac{\partial\psi}{\partial M_1}(\gamma_1, \frac{L_\mathrm{X}}{\lambda_c}, M_2, A, \lambda_c)\right], \frac{L_\mathrm{X}}{M_2}<\lambda_c<\frac{L_\mathrm{X}}{M_1}\\
\frac{\partial\psi}{\partial\lambda_c}(\gamma_1, M_1, M_2, A, \lambda_c), \lambda_c>\frac{L_\mathrm{X}}{M_1}
\end{cases}
\end{align}
\begin{align}
\frac{\partial\psi_\mathrm{DP}}{\partial\gamma_1}=
\begin{cases}
0, \lambda_c<\frac{L_\mathrm{X}}{M_2}\\
\frac{\partial\psi}{\partial\gamma}(\gamma_1, \frac{L_\mathrm{X}}{\lambda_c}, M_2, A, \lambda_c), \frac{L_\mathrm{X}}{M_2}<\lambda_c<\frac{L_\mathrm{X}}{M_1}\\
\frac{\partial\psi}{\partial\gamma}(\gamma_1, M_1, M_2, A, \lambda_c), \lambda_c>\frac{L_\mathrm{X}}{M_1}
\end{cases}
\end{align}
\begin{align}
\frac{\partial\psi_\mathrm{DP}}{\partial\gamma_2}=
\begin{cases}
\frac{\partial\psi}{\partial\gamma}(\gamma_2, M_1, M_2, A, \lambda_c), \lambda_c<\frac{L_\mathrm{X}}{M_2}\\
\frac{\partial\psi}{\partial\gamma}(\gamma_2, M_1, \frac{L_\mathrm{X}}{\lambda_c}, A, \lambda_c), \frac{L_\mathrm{X}}{M_2}<\lambda_c<\frac{L_\mathrm{X}}{M_1}\\
0, \lambda_c>\frac{L_\mathrm{X}}{M_1}
\end{cases}
\end{align}
Based on Equation~\ref{eq: phiL_mdl_psiDP}, we have
\begin{align}
\frac{\partial\phi_{L, \mathrm{mdl}}}{\partial X_{ij}}&(L_\mathrm{X}, z)=\delta_{jj_z}\times\nonumber\\
&\frac{\partial\psi_\mathrm{DP}}{\partial X}(A_{ij_z}, \lambda_{c,ij_z}, \gamma_{1,ij_z}, \gamma_{2,ij_z}, M_{LB, i}, M_{LB, i+1}; L_\mathrm{X}),
\end{align}
where $\delta_{jj_z}=0~(1)$ if $j\neq j_z~(j=j_z)$, and $X$ denotes each one of $(A, \lambda_c, \gamma_1, \gamma_2)$. The partial derivatives of $\ln\mathcal{L}_\mathrm{SMF-XLF}$ in Equation~\ref{eq: lnlike_xlf} are
\begin{align}
\frac{\partial\ln\mathcal{L}_\mathrm{SMF-XLF}}{\partial X_{ij}}=\sum_kn_k^\mathrm{XLF}\left(\frac{1}{\phi_{L, \mathrm{mdl}, k}}-\frac{1}{\phi_{L, \mathrm{obs}, k}}\right)\frac{\partial\phi_{L, \mathrm{mdl}}}{\partial X_{ij}}(L_{\mathrm{X}, k}, z_k).
\end{align}
The gradient of the posterior in Equation~\ref{eq: lnpost_smf_xlf} is
\begin{align}
\nabla\ln\mathcal{P}&=\sum_\mathrm{field}\nabla\ln\mathcal{L}+\nabla\ln\mathcal{L}_\mathrm{SMF-XLF}+\nabla\ln\pi_\mathrm{cont},
\end{align}
where $\nabla\ln\mathcal{L}$ and $\nabla\ln\pi_\mathrm{cont}$ were presented in Appendix~\ref{append: gradient}.

\section{Results without eFEDS}
\label{append: efeds}
\begin{figure*}
\centering
\resizebox{\hsize}{!}{\includegraphics{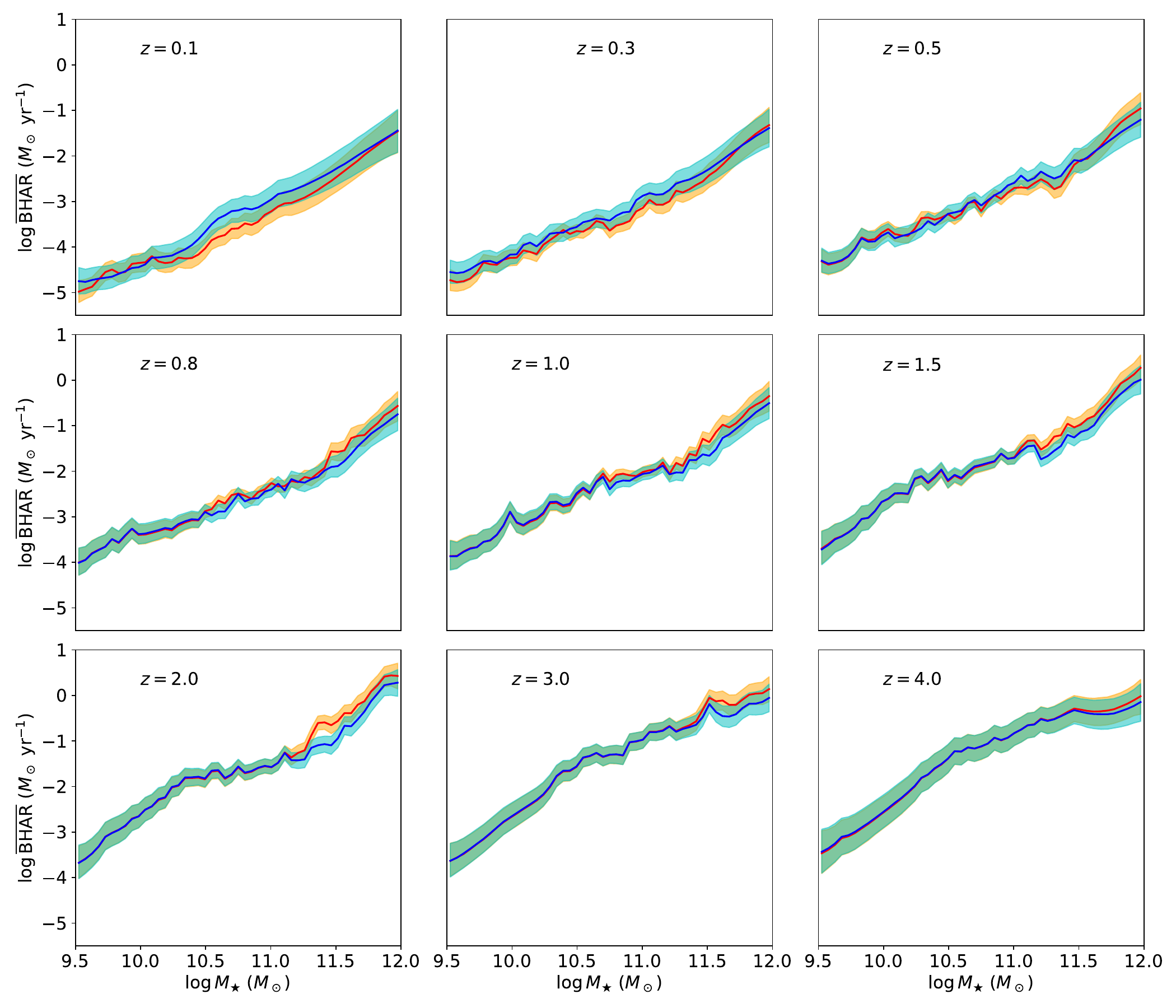}}
\caption{Comparison between $\overline{\mathrm{BHAR}}$ with eFEDS included (red) and excluded (blue) in the fitting. The shaded regions represent $1\sigma$ uncertainty ranges. The red curves are similar to the blue ones, and the red uncertainties are smaller than the blue ones in certain regimes, indicating that eFEDS does not cause systematic biases and helps constrain $\overline{\mathrm{BHAR}}$.}
\label{fig_bharcompare_efeds}
\end{figure*}

eFEDS is primarily observed through soft \mbox{X-rays} below 2~keV, which are more prone to obscuration compared to our other fields. To examine if our results are biased by this effect, we try excluding eFEDS in this appendix, and the corresponding $\overline{\mathrm{BHAR}}$ results are shown in Figure~\ref{fig_bharcompare_efeds}. There is not any material systematic difference in the median $\overline{\mathrm{BHAR}}$ after excluding eFEDS, and the uncertainty becomes larger in certain parameter ranges; e.g., the difference in width of the shaded regions in Figure~\ref{fig_bharcompare_efeds} is apparent at $M_\star\approx10^{10.8}~M_\odot$ and $z=0.5$. The uncertainties generally grow by no more than 60\%. Therefore, no strong systematic biases are introduced by eFEDS, and eFEDS also helps constrain $\overline{\mathrm{BHAR}}$. This verifies that the absorption effects have been appropriately considered, as detailed in Section~\ref{sec: Pdet}. Besides, given that the LSST DDFs already cover $12.6~\mathrm{deg^2}$ with sensitive HB data, eFEDS provides useful constraints but is not fully dominant.

\section{Impact of AGN-dominated Sources}
\label{append: blagn}
It is generally more challenging to reliably measure $M_\star$ from the galaxy component for sources with SEDs dominated by the AGN component. We assess if the less-reliable $M_\star$ measurements for such sources have strong impact on our $\overline{\mathrm{BHAR}}$ results. It has been shown that the CANDELS fields are largely free from this potential issue \citep{Aird18, Yang18a} due to their small solid angles, superb multi-wavelength coverage, and deep \mbox{X-ray} surveys. For the LSST DDFs and eFEDS, their \mbox{X-ray} surveys are wider and shallower, and thus a larger fraction of the detected AGNs are luminous and may dominate the SEDs. We thus primarily focus on the AGN-dominated sources in the LSST DDFs and eFEDS.\par
$M_\star$ is largely constrained by the rest-frame near-infrared (NIR) data because the old-star emission peaks in the NIR. For the purpose of assessing the $M_\star$ measurements, we define a source to be ``AGN-dominated'' if its AGN component contributes $>50\%$ of the rest-frame $1~\mu\mathrm{m}$ light, as measured from its decomposed SED. A similar definition was also adopted in \citet{Aird18}. About $10-15\%$ of our AGNs are classified as AGN-dominated. Note that this definition significantly overlaps but is not the same as the broad-line AGN definition. In a general sense, broad-line AGNs are sources with strong AGN signatures (e.g., spectroscopically detected broad emission lines) in the optical. However, a large fraction of broad-line AGNs are not necessarily AGN-dominated in the NIR because the galaxy emission usually reaches a peak while the AGN emission reaches a valley in the NIR. We found that around half of the broad-line AGNs in \citet{Ni21} are classified as AGN-dominated under our definition, and the non-AGN-dominated ones indeed generally have lower $L_\mathrm{X}$. We adopt our current definition because it is simpler and also more physically related to the $M_\star$ measurement.\par
\begin{figure*}
\centering
\resizebox{\hsize}{!}{\includegraphics{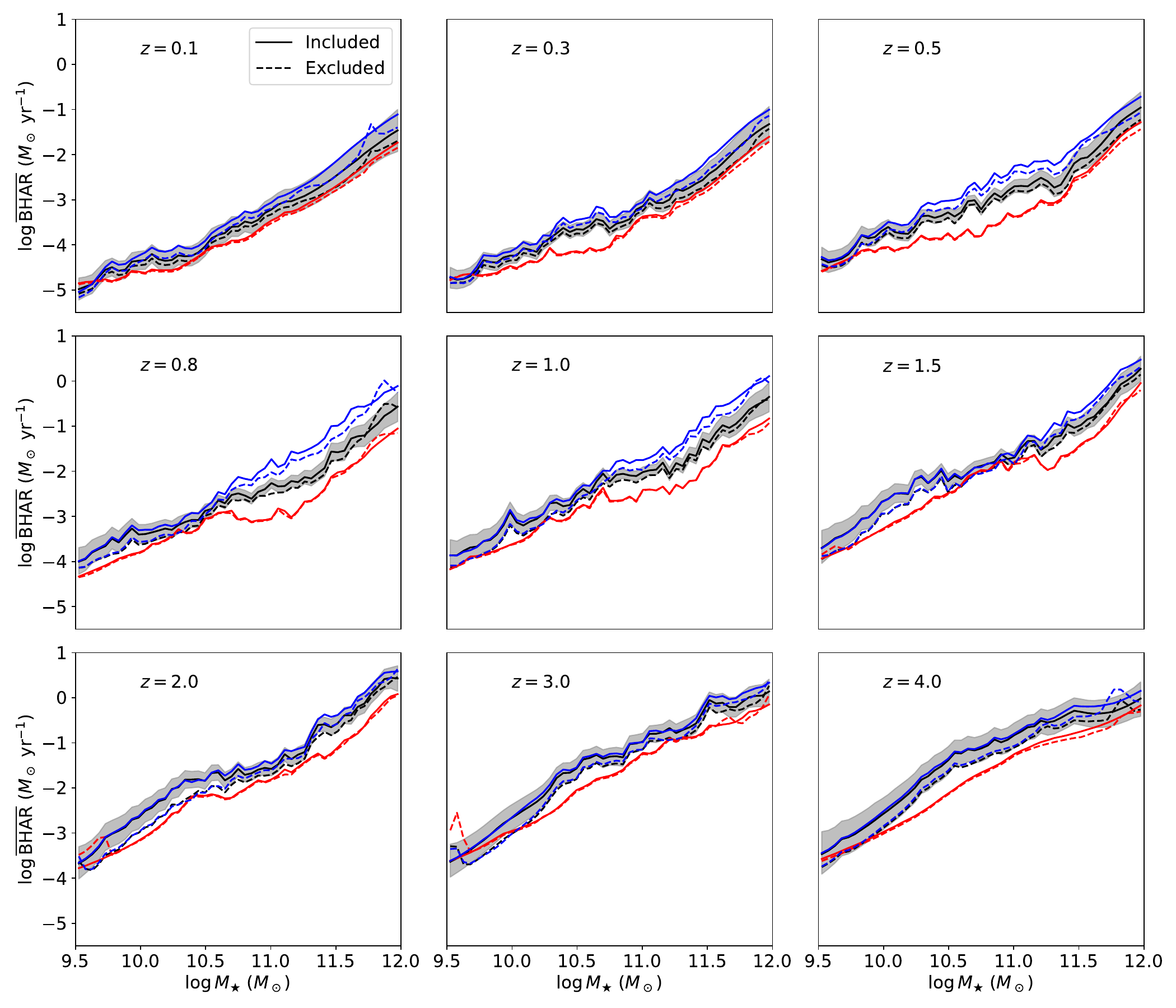}}
\caption{Comparison between $\overline{\mathrm{BHAR}}$ with AGN-dominated sources included (solid curves) and excluded (dashed curves) in the fitting. Black, blue, and red curves represent the whole population, star-forming galaxies, and quiescent galaxies, respectively. The grey shaded regions are $1\sigma$ uncertainty ranges of the black solid curves. The solid and dashed $\overline{\mathrm{BHAR}}$ curves are generally consistent within $1\sigma$ uncertainties, indicating that AGN-dominated sources do not cause material biases to our results.}
\label{fig_bharcompare_blagn}
\end{figure*}

We remove AGN-dominated sources in the LSST DDFs and eFEDS and measure $\overline{\mathrm{BHAR}}$ again following Section~\ref{sec: bhar}. We further estimate the AGN number-density maps in the $(M_\star, z)$ plane using kernel density estimations before and after excluding these AGN-dominated sources and apply the number-density ratio as a function of $(M_\star, z)$ as a correction of $\overline{\mathrm{BHAR}}$ to account for the fact that fewer AGNs are included after removing AGN-dominated sources. These procedures are conducted for the whole population as well as star-forming and quiescent galaxies. We compare $\overline{\mathrm{BHAR}}$ with the original ones in Figure~\ref{fig_bharcompare_blagn}. The quiescent curves almost do not change after removing AGN-dominated sources, while the whole-population and star-forming $\overline{\mathrm{BHAR}}$ become slightly smaller. The difference at high redshift is slightly larger than that at low redshift because high-$z$ sources need higher $L_\mathrm{X}$ to be detected in the \mbox{X-ray} and are hence more likely to be AGN-dominated, but the difference is still generally no more than the $1\sigma$ uncertainties. Besides, our number-based correction underestimates the real loss of accretion power because AGN-dominated sources, by construction, tend to have higher $\lambda$ than the remaining ones. The difference in $\overline{\mathrm{BHAR}}$ should be even smaller. Therefore, the relatively larger $M_\star$ uncertainties of AGN-dominated sources are not expected to cause material biases to our $\overline{\mathrm{BHAR}}$.

\bibliography{citations}

\end{document}